\documentclass[%
 reprint,
superscriptaddress,
 amsmath,amssymb,
 aps,
]{revtex4-2}

\usepackage{graphicx}
\usepackage{dcolumn}
\usepackage{bm}
\usepackage{braket}
\usepackage{appendix}
\usepackage{subfigure}
\usepackage{hyperref}
\usepackage{color}
\usepackage{xcolor}
\usepackage{amsmath}
\usepackage{nccmath}
\usepackage[english]{babel}
\usepackage{svg}

\begin{document}

\preprint{APS/123-QED}

\title{Thermal equilibrium spin torque: Near-field radiative angular momentum transfer in magneto-optical media}
\author{Xingyu Gao}
\affiliation{
	Department of Physics and Astronomy, Purdue University, West Lafayette, Indiana 47907, USA
}
\author{Chinmay Khandekar}
 \affiliation{%
 	Birck Nanotechnology Center, School of Electrical and Computer Engineering, College of Engineering, Purdue University, West Lafayette,
 	IN 47907, United States of America
 }
\author{Zubin Jacob}%
 \email{zjacob@purdue.edu}
\affiliation{%
 Birck Nanotechnology Center, School of Electrical and Computer Engineering, College of Engineering, Purdue University, West Lafayette,
 IN 47907, United States of America
}%
\author{Tongcang Li}%
\email{tcli@purdue.edu}
\affiliation{
	Department of Physics and Astronomy, Purdue University, West Lafayette, Indiana 47907, USA
}
\affiliation{%
	Birck Nanotechnology Center, School of Electrical and Computer Engineering, College of Engineering, Purdue University, West Lafayette,
	IN 47907, United States of America
}
\affiliation{
	Purdue Quantum Science and Engineering Institute, Purdue University, West Lafayette, Indiana 47907, USA
}

\date{\today}

\begin{abstract}
Spin and orbital angular momentum of light plays a central role in quantum nanophotonics as well as topological electrodynamics. Here, we show that the thermal radiation from finite-size bodies comprising of nonreciprocal magneto-optical materials can exert a spin torque even in global thermal equilibrium. Moving beyond the paradigm of near-field heat transfer, we calculate near-field radiative angular momentum transfer between finite-sized nonreciprocal objects by combining Rytov's fluctuational electrodynamics with the theory of optical angular momentum. We prove that a single magneto-optical cubic particle in non-equilibrium with its surroundings experiences a torque in the presence of an applied magnetic field (T-symmetry breaking). Furthermore, even in global thermal equilibrium, two particles with misaligned gyrotropy axes experience equal magnitude torques with opposite signs which tend to align their gyrotropy axes parallel to each other. Our results are universally applicable to semiconductors like InSb (magneto-plasmas) as well as Weyl semi-metals which exhibit the anomalous Hall effect (gyrotropic) at infrared frequencies. Our work paves the way towards near-field angular momentum transfer mediated by thermal fluctuations for nanoscale devices.

\end{abstract}

\maketitle


\section{Introduction}\label{Sec1:Intro}
Nanoscale radiative transfer plays an important role in a wide range of scientific and engineering disciplines. It has a variety of promising applications including energy conversion \cite{bernardi2015impacts,tervo2018near}, thermal rectification \cite{van2012tuning,ghanekar2016high},  near-field spectroscopy \cite{babuty2013blackbody,o2014spectral}, near-field super-Planckian emission \cite{guo2012broadband,biehs2013super,yang2018observing,mirmoosa2017super} etc. Over the past few years, most of the investigations in this field have focused on the near-field heat transfer i.e. energy transferred between various bodies in thermal non-equilibrium. It should be emphasized that in global equilibrium there is no net flow of energy in these systems \cite{liu2014near,liu2013near,biehs2011nanoscale}. Our goal in this paper is to explore concepts beyond energy i.e. angular momentum which can be exchanged/transferred between bodies even in global thermal equilibrium. Thus our result is similar to the Casimir torque \cite{somers2018measurement} obtained using birefringent crystals but using a different underlying mechanism based on nonreciprocal materials.

Thermal spin photonics is an emerging research area that combines the thermal radiation and the spin angular momentum (SAM) of light \cite{ott2018circular,khandekar2019circularly,bliokh2015spin,ott2020thermal}. Non-reciprocal materials such as semiconductors in external magnetic fields \cite{armelles2013magnetoplasmonics,sengupta2020electron,ishimaru2017electromagnetic} and Weyl semimetals \cite{kotov2018giant} are of great interest in the context of thermal spin photonics as they break the time reversal symmetry and lead to many interesting effects \cite{zhu2014near,latella2017giant,ben2016photon,herz2019green,moncada2015magnetic,van2018quantum}. Recent work showed that the thermal radiation of a non-reciprocal medium carries angular momentum (AM) \cite{khan2019spinning,khandekar2019thermal,ott2018circular,maghrebi2019fluctuation}. This can further induce a torque on the media.  There also exist new spin-resolved thermal radiation laws applicable for nonreciprocal bianisotropic media, which extend known Kirchhoff's laws otherwise valid only for reciprocal media \cite{khandekar2019universal}. However, these investigations are limited to planar geometries and dipolar particles. More complex structured surfaces or finite-size bodies have remained unexplored because of computational difficulty. Recently, significant progress has been made on numerical approaches in the context of near-field heat transfer with nontrivial geometries. Those computational tools include the scattering matrix \cite{kruger2011nonequilibrium,kruger2012trace}, boundary-element methods \cite{rodriguez2012fluctuating,rodriguez2013fluctuating}, volume-integral-equation methods \cite{polimeridis2015fluctuating}, and the thermal discrete dipole approximation method (TDDA) \cite{edalatpour2014thermal,edalatpour2015convergence,ekeroth2017thermal}.  However, so far none of them has been used for analyzing radiative angular momentum transfer in non-reciprocal bodies. Here, we primarily aim to explore thermal AM radiation in the near field and far field of finite-size nonreciprocal objects. 

\begin{figure*}[t]
    \centering
	\includegraphics[width=0.7\textwidth]{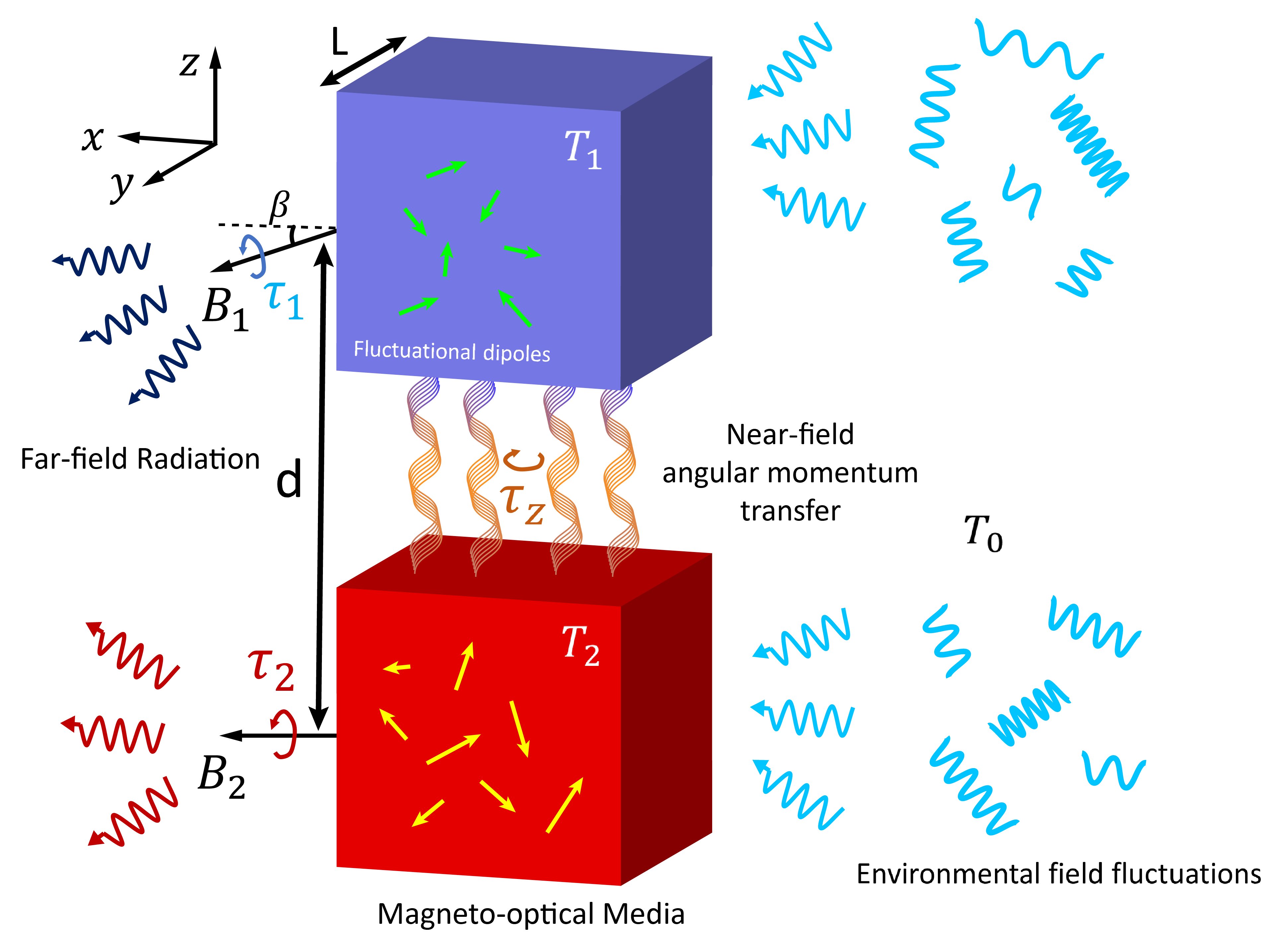}
	\caption{(Color online) Schematic of the radiative AM transfer in a two-body MO system in the presence of external magnetic fields. Two cubes are separated by a distance d (center to center distance) and they have the same size L. $T_1$, $T_2$ and $T_0$ denote the temperatures of the top cube, bottom cube and environment, respectively. } \label{fig:1}
\end{figure*}

We explore the AM-resolved thermal radiation features in a system comprising magneto-optical bodies in external magnetic fields (as shown in Figure \ref{fig:1}). Magneto-optical (MO) media, such as doped indium antimonide (InSb), become gyroelectric objects in the presence of magnetic fields and hence exhibit stable non-reciprocity that can be controlled by changing the magnitudes and the directions of the magnetic fields \cite{moncada2015magnetic,ben2016photon}. We find that the thermal radiation from a single gyroelectric body carries a net AM flux directed along its gyrotropy axis (or the direction of externally applied magnetic field) when the system is out of equilibrium with vacuum. Consequently, the particle experiences a net torque along the opposite direction based on angular momentum conservation. More interestingly, we also find that for a two-body system with misaligned gyrotropy axes, AM transfer between two bodies can occur regardless of the global thermal equilibrium. The AM transfer results in a torque with the same magnitude but opposite signs on two bodies trying to align the gyrotropy axes parallel to each other. Moreover, the total torque on the combined system is zero, which indicates that there is no net AM flux transferred to the far field, satisfying the detailed balance of AM flux between the combined system and the environment at thermal equilibrium.

In this work, we utilize fluctuational electrodynamics (FE) that combines the Maxwell electromagnetic theory of angular momentum and the fluctuation-dissipation theorem (FDT) \cite{polder1971theory}. In particular, our numerical approach is generalized from the TDDA method by a recent work studying the thermal emission in MO systems \cite{ekeroth2017thermal}. In that work, the authors proved the validity of the TDDA method for optically anisotropic systems that can be described by an arbitrary electric permittivity tensor (with $\mu=1$). We extend the TDDA approach to describe the thermal AM flux in the near-field and far-field of finite nonreciprocal bodies. Moreover, we apply our TDDA approach for exploring the near-field and far-field transfer of angular momentum in MO systems. We elucidate the origin of the thermal-fluctuations-induced torque by computing the angular momentum flux across different planes. 

We also demonstrate the nontrivial role of non-reciprocity in causing an equilibrium torque in a system which is at global thermal equilibrium. Such an equilibrium torque is non-intuitive but important since it opens a degree of freedom for directional radiative AM transfer. Experimental demonstration of such an effect requires misaligned magnetic fields which can be generated using spatial gradients of two-dimensional magnetic fields on nanoscale objects. Our work provides a way to explore many interesting effects in the context of thermal AM in nonreciprocal systems. As the dimensions of our system under consideration are smaller than  the wavelengths of thermal radiation, the contribution of angular momentum is dominated by spin as opposed to orbital angular momentum. This observation is consistent with recent experiments in ion traps where orbital angular momentum has been shown to couple only to quadrupolar optical transitions, not dipolar optical transitions \cite{schmiegelow2012light,schmiegelow2016transfer}. To emphasize the analogy with spin transfer torque in nanoelectronics \cite{urazhdin2014nanomagnonic,ralph2008spin}, we term our phenomenon as thermal photonic spin transfer torque. We also note that our work is applicable not only to magnetized-plasmas like InSb but also to Weyl semi-metal particles which show the anomalous Hall effect (non-reciprocity) without an applied magnetic field \cite{guo2020radiative,zhao2020axion,tsurimaki2020large}.

The rest of the paper will be organized as follows. In Sec. \ref{Sec2:Theory}, we show our theoretical formalism of the TDDA method to describe the radiative AM transfer in the near-field and far-field of MO objects. In Sec. \ref{Sec3:Results}, we discuss our numerical results obtained by our TDDA approach. We separate our discussion into the single-cube case and two-cube case. For each case, we consider both the thermal equilibrium and non-equilibrium conditions. In Sec. \ref{Sec:Conclusion}, we summarize our observations in Sec. \ref{Sec3:Results} and conclude the paper with some additional remarks.

\section{\label{Sec2:Theory} Angular momentum in fluctuational electrodynamics}

\subsection{\label{subsec2_1:Angular} Radiative angular momentum flux} 
We consider a single- or two-cube system made of nonreciprocal materials shown in Figure \ref{fig:1}. We focus on the AM-resolved thermal radiation on the vacuum side of the systems. For this purpose, momentum and AM will be studied in the context of the thermal radiation. These quantities are important because they are conserved and thus follow the conservation laws. First, the radiative momentum at the observation point can be quantified by Poynting flux $\mathbf{P}$ and the momentum flux density $\mathbf{\Sigma}$ \cite{barnett2001optical}:
\begin{equation}\label{Eq:Poynting}
\left\langle\mathbf{P} \right\rangle=\left\langle \mathbf{E} \times \mathbf{H}\right\rangle,
\end{equation}
\begin{equation}\label{Eq:MaxwellStress}
\begin{aligned}
\left\langle \mathbf{\Sigma}\right\rangle=&-\epsilon_0\left\langle\mathbf{E} \otimes \mathbf{E}\right\rangle-\mu_0\left\langle\mathbf{H} \otimes \mathbf{H}\right\rangle\\
&+\frac{1}{2}\epsilon_0\operatorname{Tr}\left\{\left\langle\mathbf{E} \otimes \mathbf{E}\right\rangle\right\}\mathbb{I}+\frac{1}{2}\mu_0\operatorname{Tr}\left\{\left\langle\mathbf{H} \otimes \mathbf{H}\right\rangle\right\} \mathbb{I}
\end{aligned}
\end{equation}
where $\otimes$ denotes the outer product of two vectors, and $\left\langle ...\right\rangle$ denotes the thermodynamic ensemble average. All the quantities in Eq. \ref{Eq:Poynting} and Eq. \ref{Eq:MaxwellStress} are dependent on the position $\mathbf{r}$ and time $t$. The radiation momentum transfer leads to a force $\mathbf{F}$ on the objects that can be obtained by the conservation law of momentum: 

\begin{equation}\label{Eq:force}
\int_{S} \left\langle \Sigma_{i j}(\mathbf{r},t) \right\rangle \mathrm{d} S_{j}=-\frac{\partial}{\partial t} \left (\int_{V} \left\langle\mathbf{P}(\mathbf{r},t) \right\rangle \mathrm{d} V \right )_{i} = -\left\langle F_{i} \right\rangle.
\end{equation}

Similarly, the radiation AM can be quantified by AM density $\textbf{J}$ and AM flux density $\textbf{M}$ \cite{barnett2001optical}:
\begin{align}
\left\langle\mathbf{J}(\mathbf{r},t) \right\rangle=\mathbf{r}\times \left\langle\mathbf{E}(\mathbf{r},t) \times \mathbf{H}(\mathbf{r},t)\right\rangle
\end{align}
\noindent
\begin{equation}\label{Eq:AngularDensity}
\left\langle\mathbf{M}(\mathbf{r},t) \right \rangle=\mathbf{r}\times \left\langle \mathbf{\Sigma}(\mathbf{r},t)\right\rangle
\end{equation}
which is given by the cross product of the position with the Maxwell stress tenor. They also satisfy the continuity equation
\begin{equation}\label{Eq:Continuity}
\frac{\partial}{\partial t}J_i+\frac{\partial}{\partial x_l}M_{li}=0.
\end{equation}

A nonreciprocal medium can lead to a nonzero radiation AM flux and thus results in a torque $\tau$ on objects. This torque can be obtained by the conservation law of AM, which is the integral form of the continuity equation in Eq. \ref{Eq:Continuity}:
\begin{equation}\label{Eq:torque}
\int_{S} \left\langle M_{i j}(\mathbf{r},t) \right\rangle \mathrm{d} S_{j}=-\frac{\partial}{\partial t} \left (\int_{V} \left\langle\mathbf{J}(\mathbf{r},t) \right\rangle \mathrm{d} V \right )_{i} = -\left\langle \tau_{i} \right\rangle.
\end{equation}

All the quantities throughout the paper are described in SI units. Next, we express the physical quantities in terms of their Fourier transforms, such as

\begin{align}
\mathbf{E}(t)&=\int_{-\infty}^{\infty} \frac{d \omega}{2 \pi} \mathbf{E}(\omega) e^{-i \omega t},\\ 
\mathbf{H}(t)&=\int_{-\infty}^{\infty} \frac{d \omega}{2 \pi} \mathbf{H}(\omega) e^{-i \omega t}.
\end{align}
for electromagnetic field, with similar notation for other quantities. Then above quantities $\mathbf{Q}=\left\{\mathbf{\Sigma},\mathbf{M},\mathbf{F},\tau\right\}$ are to be integrated over frequency to obtain the total flux/force/torque as $\mathbf{Q}=\int_{-\infty}^{\infty} \frac{d \omega}{2 \pi} \mathbf{Q}(\omega) e^{-i \omega t}$.
The electromagnetic field correlations required for calculating densities and flux rates above are obtained from the FDT. We can separate the contributions into two parts: the first one accounts for the fluctuations of particle dipole moments that will further induce electromagnetic field; the second part involves environmental field fluctuations. The FDT in our case gives \cite{messina2013fluctuation}

\begin{equation} \label{Eq:FDT1}
\begin{aligned}
\left\langle p_{\mathrm{f}, i}(\omega) p_{\mathrm{f}, j}^*\left(\omega^{\prime}\right)\right\rangle=& 2 \pi \hbar\epsilon_0 \delta\left(\omega-\omega^{\prime}\right) \\
& \times \operatorname{Im}\left\{\alpha_{i j}(\omega)\right\}\left(1+2n_{p}(\omega)\right)
\end{aligned}
\end{equation}
for the electric dipole fluctuations and
\begin{equation} \label{Eq:FDT2}
\begin{aligned}
\left\langle E_{\mathrm{f}, i}(\mathbf{r}, \omega) E_{\mathrm{f}, j}^*\left(\mathbf{r}^{\prime}, \omega^{\prime}\right)\right\rangle=& 2 \pi \frac{\hbar k_0^2}{\epsilon_0} \delta\left(\omega-\omega^{\prime}\right) \\
& \times \operatorname{Im}\left\{G_{EE,i j}\left(\mathbf{r}, \mathbf{r}^{\prime}\right)\right\}\left(1+2n_{0}(\omega)\right)
\end{aligned}
\end{equation}
for the environmental electric-field fluctuations, where $k_0=\omega/c$ is the magnitude of the vacuum wave vector. Here $\hat{\alpha}$ is the polarizability of the objects and $\mathbf{G}_{EE}(\mathbf{r},\mathbf{r}^{\prime})$ is the free space electric-electric dyadic Green's tensor \cite{messina2013fluctuation}. The temperature of the particle $T_p$ and the vacuum $T_0$ enter these expressions through the Bose-Einstein distribution
$
n_{l}(\omega)=1/(e^{\hbar \omega / k_{B} T_{l}}-1).
$

\subsection{Numerical Approach: Thermal Discrete Dipole Approximation}
To numerically compute the physical quantities introduced in Eq. \ref{Eq:AngularDensity} and Eq. \ref{Eq:torque}, we extend the TDDA approach for MO objects based on Ref. \cite{ekeroth2017thermal}. Here we summarize the important equations we have developed. More details can be found in Appendix \ref{Append:TDDA}. 

Considering a two-body system interacting with a thermal bath, we use a collection $N_p$ (for object p) electric point dipoles to describe the system. Each dipole is characterized by a volume $V_{i,p}$ and a polarizability tensor $\hat{\alpha}_{i,p}$, where p=1,2 denotes the body that the dipole belongs to and i=1,2,...,$N_p$ indicates the i-th subvolume in that object. We group the electric dipoles and electric fields inside bodies in a compact form:
\begin{equation}
\begin{array}{l}
\overline{\mathbf{P}}=\left(\begin{array}{c}
\overline{\mathbf{P}}_{1} \\
\overline{\mathbf{P}}_{2}
\end{array}\right) ; \quad \overline{\mathbf{P}}_{1}=\left(\begin{array}{c}
\mathbf{p}_{1,1} \\
\vdots \\
\mathbf{p}_{N_{1}, 1}
\end{array}\right), \overline{\mathbf{P}}_{2}=\left(\begin{array}{c}
\mathbf{p}_{1,2} \\
\vdots \\
\mathbf{p}_{N_{2}, 2}
\end{array}\right) \\
\overline{\mathbf{E}}=\left(\begin{array}{c}
\overline{\mathbf{E}}_{1} \\
\overline{\mathbf{E}}_{2}
\end{array}\right) ; \quad \overline{\mathbf{E}}_{1}=\left(\begin{array}{c}
\mathbf{E}_{1,1} \\
\vdots \\
\mathbf{E}_{N_{1}, 1}
\end{array}\right), \overline{\mathbf{E}}_{2}=\left(\begin{array}{c}
\mathbf{E}_{1,2} \\
\vdots \\
\mathbf{E}_{N_{2}, 2}
\end{array}\right)
\end{array}.
\end{equation}
The notation $\bar{\mathbf{C}}$ ($\mathbf{C}=\{\mathbf{p},\mathbf{E},...\}$) indicates that the dipoles or the field are inside bodies and in the following discussion, we denote $\mathbf{C}$ without the overhead bar as the quantities in vacuum. 

To obtain the total radiative AM flux in the near field or far field (obsevation point), we need to compute the statistical average of the AM flux density introduced in Eq. (\ref{Eq:AngularDensity}). Making use of the Fourier transforms, the average of the AM flux density can be expressed as 
\begin{equation} \label{Eq:TDDA_tau}
\begin{aligned}
\left\langle \mathbf{M}(\mathbf{r})\right\rangle=& -2 \int_{0}^{\infty} \frac{d \omega}{2 \pi} \int_{-\infty}^{\infty} \frac{d \omega^{\prime}}{2 \pi}\mathbf{r}\times\\
&\operatorname{Re}[ \epsilon_0\left\langle\mathbf{E}(\mathbf{r}, \omega) \otimes \mathbf{E}^*(\mathbf{r}, \omega^{\prime})\right\rangle e^{-i\left(\omega-\omega^{\prime}\right) t}\\
&+\mu_0\left\langle\mathbf{H}(\mathbf{r}, \omega) \otimes \mathbf{H}^*(\mathbf{r}, \omega^{\prime})\right\rangle e^{-i\left(\omega-\omega^{\prime}\right) t}\\
&-\frac{1}{2}\epsilon_0\operatorname{Tr}\left\{\left\langle\mathbf{E}(\mathbf{r}, \omega) \otimes \mathbf{E}^*(\mathbf{r}, \omega^{\prime})\right\rangle e^{-i\left(\omega-\omega^{\prime}\right) t}\right\}\mathbb{I}\\
&-\frac{1}{2}\mu_0\operatorname{Tr}\left\{\left\langle\mathbf{H}(\mathbf{r}, \omega) \otimes \mathbf{H}^*(\mathbf{r}, \omega^{\prime})\right\rangle e^{-i\left(\omega-\omega^{\prime}\right) t}\right\}\mathbb{I}].
\end{aligned}
\end{equation}
Using FDT from Eq. \ref{Eq:FDT1} and Eq. \ref{Eq:FDT2}, the above expression can be reduced to the integration of the terms containing $\left\langle\mathbf{E}(\mathbf{r}, \omega) \otimes \mathbf{E}^*(\mathbf{r}, \omega)\right\rangle$ and $\left\langle\mathbf{H}(\mathbf{r}, \omega) \otimes \mathbf{H}^*(\mathbf{r}, \omega)\right\rangle$. As we discussed in the last section, Eq. \ref{Eq:FDT1} and Eq. \ref{Eq:FDT2} introduce two different sources that contribute to the total field correlations. The first one is the fluctuating particle dipoles determined by the temperatures of bodies. In the TDDA approach, it can be written as:
\begin{equation}
\left\langle\bar{\mathbf{P}}_{\mathrm{f}}(\omega) \bar{\mathbf{P}}_{\mathrm{f}}^{\dagger}\left(\omega^{\prime}\right)\right\rangle=2 \pi\hbar \epsilon_{0}  \delta\left(\omega-\omega^{\prime}\right)\left[\mathbb{I}+2 \hat{n}_{\mathrm{B}}\left(\omega, T_1, T_2\right)\right] \hat{\chi},
\end{equation}
where $T_1$ and $T_2$ are the temperatures of the two objects, and $ \hat{n}_{\mathrm{B}}\left(\omega, T_1, T_2\right)$ is a diagonal tensor with $3N$ elements given by the Bose-Einstein distribution. $\hat{\chi}$ is a tensor combining the imaginary part of the polarization tensor and the radiative correction \cite{manjavacas2012radiative,messina2013fluctuation}.
Using the general TDDA equations and some algebraic manipulations, we obtain the following expressions for the correlations of electric and magnetic fields at the observation point outside bodies:
\begin{equation} \label{Eq:E_cor_p}
\left\langle\mathbf{E}(\mathbf{r}, \omega) \otimes \mathbf{E}^*(\mathbf{r}, \omega)\right\rangle=\frac{k_0^4}{\epsilon_{0}^2}\mathbf{G}_{EE}\bar{\mathbf{T}}^{-1}\left\langle\bar{\mathbf{P}}_{\mathrm{f}} \bar{\mathbf{P}}_{\mathrm{f}}^{\dagger}\right\rangle\bar{\mathbf{T}}^{-1\dagger}\mathbf{G}_{EE}^{\dagger},
\end{equation}
\begin{equation}  \label{Eq:H_cor_p}
\left\langle\mathbf{H}(\mathbf{r}, \omega) \otimes \mathbf{H}^*(\mathbf{r}, \omega)\right\rangle=\frac{k_0^4}{\epsilon_{0}^2}\mathbf{G}_{HE}\bar{\mathbf{T}}^{-1}\left\langle\bar{\mathbf{P}}_{\mathrm{f}} \bar{\mathbf{P}}_{\mathrm{f}}^{\dagger}\right\rangle\bar{\mathbf{T}}^{-1\dagger}\mathbf{G}_{HE}^{\dagger}
\end{equation}
where $\mathbf{T}_{i j}=\delta_{i j} \mathbb{I}-\left(1-\delta_{i j}\right) k_0^{2} \bar{\alpha_{i}} \bar{\mathbb{G}}_{EE,i j}$ and $\mathbf{G}_{HE}$ is the magnetic-electric field Green's tensor (Eq. \ref{G_HE}). More details about the derivation are shown in Appendix \ref{subsec2_2_1:pfl}. 

The second source is the environmental field fluctuation. The fluctuational electric fields in the vacuum can induce electric dipoles on the objects and then generate electromagnetic fields at the observation point. In this case, the field-field correlations at the observation point outside bodies are given by (Appendix \ref{subsec2_2_2:pf2}):
\begin{equation} \label{Eq:E_cor_E}
\begin{aligned}
\left\langle\mathbf{E}(\mathbf{r}, \omega) \otimes \mathbf{E}^*(\mathbf{r}, \omega)\right\rangle=&k_0^4\mathbf{G}_{EE}\bar{\mathbf{T}}^{-1}\bar{\alpha}\left\langle\bar{\mathbf{E}}_{\mathrm{f}} \bar{\mathbf{E}}_{\mathrm{f}}^{\dagger}\right\rangle\bar{\alpha}^\dagger\bar{\mathbf{T}}^{-1\dagger}\mathbf{G}_{EE}^{\dagger}\\
&+k_0^2\mathbf{G}_{EE}\bar{\mathbf{T}}^{-1}\bar{\alpha}\left\langle\bar{\mathbf{E}}_{\mathrm{f}} \mathbf{E}_{\mathrm{f}}^{\dagger}\right\rangle\\
&+k_0^2\left\langle\mathbf{E}_{\mathrm{f}} \bar{\mathbf{E}}_{\mathrm{f}}^{\dagger}\right\rangle\bar{\alpha}^\dagger\bar{\mathbf{T}}^{-1\dagger}\mathbf{G}_{EE}^{\dagger},
\end{aligned}
\end{equation}

\begin{equation} \label{Eq:H_cor_E}
\begin{aligned}
\left\langle\mathbf{H}(\mathbf{r}, \omega) \otimes \mathbf{H}^*(\mathbf{r}, \omega)\right\rangle=&k_0^4\mathbf{G}_{HE}\bar{\mathbf{T}}^{-1}\bar{\alpha}\left\langle\bar{\mathbf{E}}_{\mathrm{f}} \bar{\mathbf{E}}_{\mathrm{f}}^{\dagger}\right\rangle\bar{\alpha}^\dagger\bar{\mathbf{T}}^{-1\dagger}\mathbf{G}_{HE}^{\dagger}\\
&+k_0^2\mathbf{G}_{HE}\bar{\mathbf{T}}^{-1}\bar{\alpha}\left\langle\bar{\mathbf{E}}_{\mathrm{f}} \mathbf{H}_{\mathrm{f}}^{\dagger}\right\rangle\\
&+k_0^2\left\langle\mathbf{H}_{\mathrm{f}} \bar{\mathbf{E}}_{\mathrm{f}}^{\dagger}\right\rangle\bar{\alpha}^\dagger\bar{\mathbf{T}}^{-1\dagger}\mathbf{G}_{HE}^{\dagger}.
\end{aligned}
\end{equation}
where 
\begin{equation}
\left\langle\bar{\mathbf{E}}_{\mathrm{f}}(\omega) \bar{\mathbf{E}}_{\mathrm{f}}^{\dagger}\left(\omega^{\prime}\right)\right\rangle=2 \pi\frac{\hbar k_0^2}{\epsilon_{0} }  \delta\left(\omega-\omega^{\prime}\right)\left(1+2 n_{\mathrm{0}}\left(\omega\right)\right) \operatorname{Im}\bar{\mathbb{G}}_{EE},
\end{equation}

\begin{equation} \label{Eq:correlat1}
\left\langle\bar{\mathbf{E}}_{\mathrm{f}}(\omega) \mathbf{E}_{\mathrm{f}}^{\dagger}\left(\omega^{\prime}\right)\right\rangle=2 \pi\frac{\hbar k_0^2}{\epsilon_{0} }  \delta\left(\omega-\omega^{\prime}\right)\left(1+2 n_{\mathrm{0}}\left(\omega\right)\right) \operatorname{Im}\mathbf{G}_{EE},
\end{equation}

\begin{equation} \label{Eq:correlat2}
\left\langle\mathbf{H}_{\mathrm{f}}(\omega) \bar{\mathbf{E}}_{\mathrm{f}}^{\dagger}\left(\omega^{\prime}\right)\right\rangle=2 \pi\frac{\hbar k_0^2}{\epsilon_{0} } \delta\left(\omega-\omega^{\prime}\right)\left(1+2 n_{\mathrm{0}}\left(\omega\right)\right) \operatorname{Im}\mathbf{G}_{HE}.
\end{equation}
are the vacuum electric-electric field correlation inside bodies, the vacuum electric-electric field correlation between the object position and observation point, and the vacuum electric-magnetic field correlation between the object position and observation point, respectively. Generally, these two sources have opposite contributions to the total far field radiations and at global thermal equilibrium, they should cancel each other to satisfy the detailed balance of the radiative AM flux.

\section{\label{Sec3:Results} Numerical Results}
In this section, we present the numerical results related to the radiative AM transfer and the induced torques discussed in the previous sections. In order to explore the role played by nonreciprocity, we consider the near-field and far-field radiation from single- and two-cube systems that are made of nonreciprocal materials. Here we choose doped InSb as an example. InSb is a MO material whose permittivity model has been well-characterized experimentally \cite{hartstein1975investigation,palik1976coupled,chochol2016magneto,chochol2017experimental}. Subjected to an external magnetic field, it shows the gyroelectric properties with gyrotropy axis along the magnetic field. The permittivity tensor in an arbitrary magnetic field takes the form of  $\overline{\bar{\varepsilon}}=\varepsilon_{\infty}\left[1+\left(\omega_{L}^{2}-\omega_{T}^{2}\right) /\left(\omega_{T}^{2}-\omega^{2}-\mathrm{i} \Gamma \omega\right)\right] \mathbb{I}_{3 \times 3}+\varepsilon_{\infty} \omega_{p}^{2}\left[\mathbb{L}_{3} \times 3(\omega)\right]^{-1}$ \cite{khandekar2019thermal}, where 
\begin{equation*}
\mathbb{L}_{3 \times 3}(\omega)=\left[\begin{array}{ccc}
-\omega^{2}-\mathrm{i} \gamma \omega & -\mathrm{i} \omega \omega_{c z} & \mathrm{i} \omega \omega_{c y} \\
\mathrm{i} \omega \omega_{c z} & -\omega^{2}-\mathrm{i} \gamma \omega & -\mathrm{i} \omega \omega_{c x} \\
-\mathrm{i} \omega \omega_{c y} & \mathrm{i} \omega \omega_{c x} & -\omega^{2}-\mathrm{i} \gamma \omega
\end{array}\right].
\end{equation*}
Here, $\epsilon_\infty$ is the high-frequency dielectric constant, $\omega_L$ is the longitudinal optical-phonon frequency, $\omega_T$ is the transverse optical-phonon frequency, and $\omega_p$ is the plasma frequency of the free carriers of density n. $\Gamma$ is the phonon damping constant, and $\gamma$ is the free-carrier damping constant. And $\omega_{ci}$ (i=x,y,z) is the cyclotron frequency given by $\omega_{ci}=qB_j/m_f$. All the parameters are taken from \cite{palik1976coupled,chochol2016magneto}, where doping density $n$ = $10^{17}$ cm$^{-3}$, $\epsilon_{\infty}$ = 15.7, $\omega_L$ = 3.62 $\times$ $10^{13}$ rad s$^{-1}$, $\omega_T$ = 3.39 $\times$ $10^{13}$ rad s$^{-1}$, $\omega_p$ = 3.14 $\times$ $10^{13}$ rad s$^{-1}$, $\Gamma$ = 5.65 $\times$ $10^{11}$ rad s$^{-1}$, $\gamma$ = 3.39 $\times$ $10^{12}$ rad s$^{-1}$, and $m_f$ = 0.022$m_e$, where $m_e$ = 9.1094 $\times$ $10^{-31}$ kg is electron mass.  

\begin{figure}[h]
	\centering
	\includegraphics[width=0.5\textwidth]{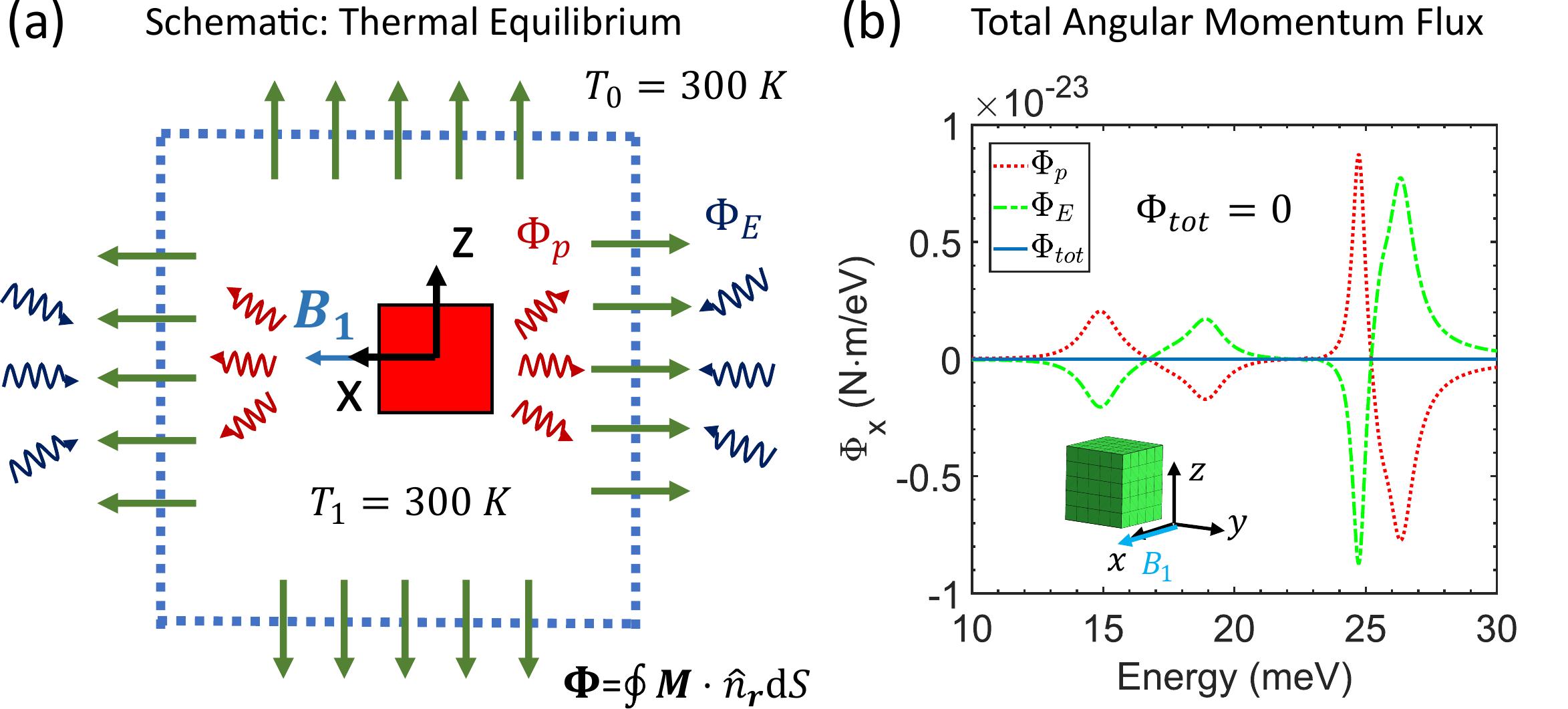}
	\caption{(Color online) (a) Schematic: Thermal radiation of the AM from a single InSb cube at thermal equilibrium ($T_1=T_2=300 K$). A 1 T external magnetic field along x direction is applied on the cube. A surface (virtual cube shown by the blue dashed line) that encloses the cube (the real physical cube is shown in red) is chosen to compute the total AM flux (green arrows) radiated to the far field. (b) The spectrum of the total radiative AM flux at thermal equilibrium. The total flux $\Phi_{tot}$ is separated into two parts: one comes from the particle dipole fluctuations (denoted as $\Phi_p$) and the other one originates from the environmental field fluctuations (denoted as $\Phi_E$). At thermal equilibrium, $\Phi_p$ and $\Phi_E$ have the same magnitude but opposite signs, resulting in a zero total flux in the far field. } \label{fig:2}
\end{figure}

\begin{figure*}[t]

	\centering
	\includegraphics[width=1\textwidth]{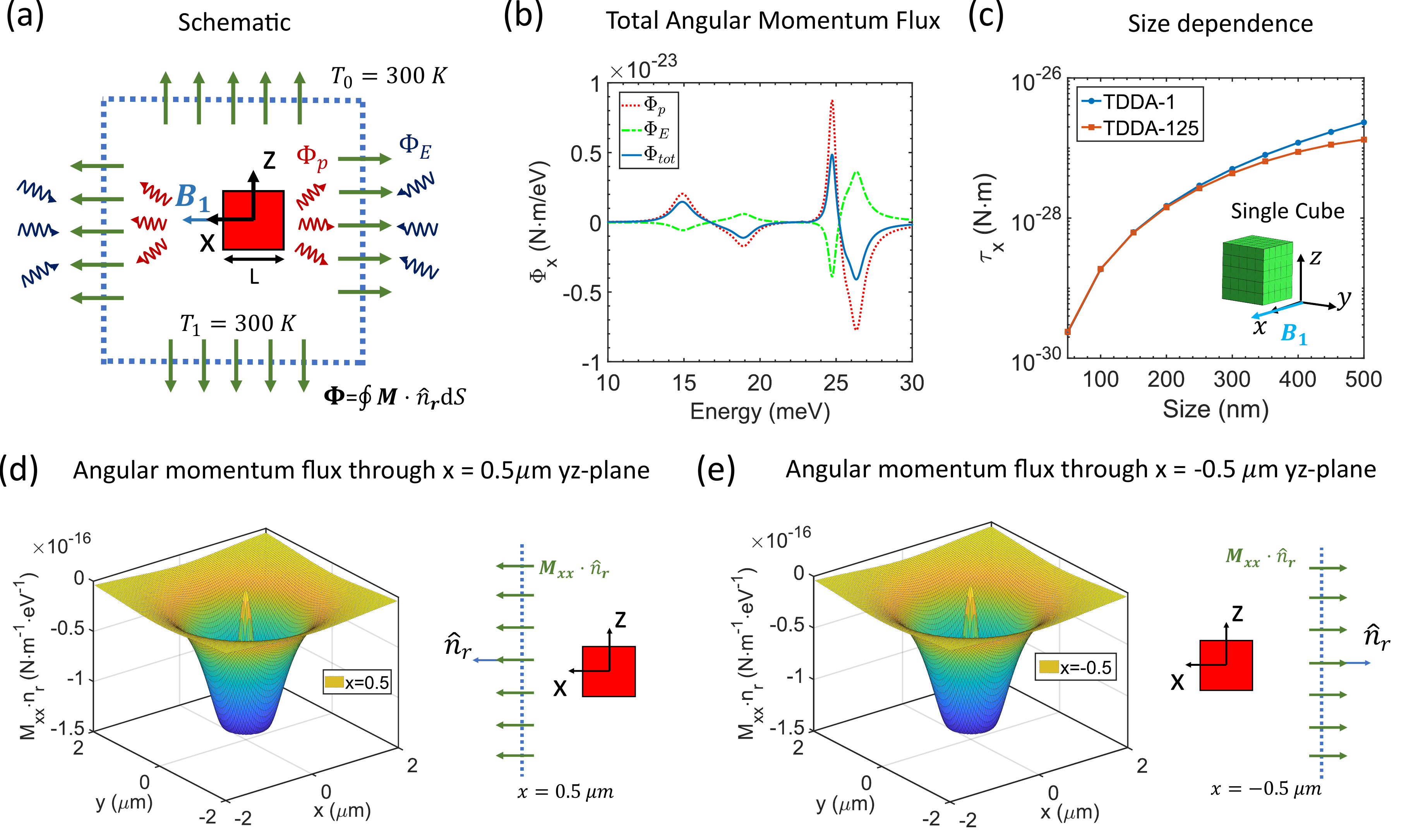}
	\caption{(Color online) Thermal radiation of the AM from a single InSb cube at thermal non-equilibrium.  (a) Schematic: A cube is in the presence of a magnetic field of 1 T along the x direction. The cube is at 300 K while the environment is at 0 K. A surface that encloses the cube is chosen to compute the AM flux radiated to the far field. (b) Spectrum of $\Phi_x$. At thermal non-equilibrium, $\Phi_p$ and $\Phi_E$ have unequal magnitudes and hence the total AM flux $\Phi_{tot}$ has a non-zero value. (c) Thermal non-equilibrium torque due to the AM radiation as a function of the cube size. The blue line with dot markers is computed by TDDA-1 (approximating the entire cube by a single dipole) and red line with square markers is obtained by using TDDA-125 (dividing each cube into 125 subvolumes and regarding each subvolume as a single dipole). (d) and (e) the spatial distribution of the AM flux through yz-planes at x = 0.5 $\mu m$ and  x = -0.5 $\mu m$, respectively. Spatial distributions are plotted at the energy 0.2475 eV in thermal non-equilibrium case. The cube is assumed to be placed at origin.} \label{fig:3}
\end{figure*}

\subsection{\label{Sec3_1:Single} Thermal AM radiation from a single cube}
As shown in figure \ref{fig:2}(a) and \ref{fig:3}(a), a single cube made of InSb is in the presence of an uniform magnetic field of 1 T along the x direction. The first issue we want to address now is the description of the thermal AM radiation from the single cube using the formalism detailed in Section \ref{Sec2:Theory}. Instead of using the SAM density $\mathbf{S}(\mathbf{r})=\frac{\epsilon_{0}}{2 \omega} \operatorname{Im}\left\langle\mathbf{E}^{*}(\mathbf{r}) \times \mathbf{E}(\mathbf{r})\right\rangle+\frac{\mu_{0}}{2 \omega} \operatorname{Im}\left\langle\mathbf{H}^{*}(\mathbf{r}) \times \mathbf{H}(\mathbf{r})\right\rangle$ that was utilized to quantify the spin component of the thermal radiation in our former work \cite{khandekar2019thermal}, here we define a tensor of angular-momentum flux  density $\mathbf{M}$ (Eq. \ref{Eq:AngularDensity}) in analogy to the well-known Maxwell stress tensor. It allows us to compute the total flux $\Phi_{tot}$ of the radiative AM in the near-field and far-field, making a better connection between the thermal radiation and the induced torques. This definition includes the spin and orbital parts of the electromagnetic field. It reveals the whole story of the AM in the radiation and will not contradict the debate about the angular-momentum separation for electromagnetic fields \cite{barnett2016natures,bliokh2014conservation,cameron2012optical}.

\begin{figure*}[t]
	\centering
	\includegraphics[width=1\textwidth]{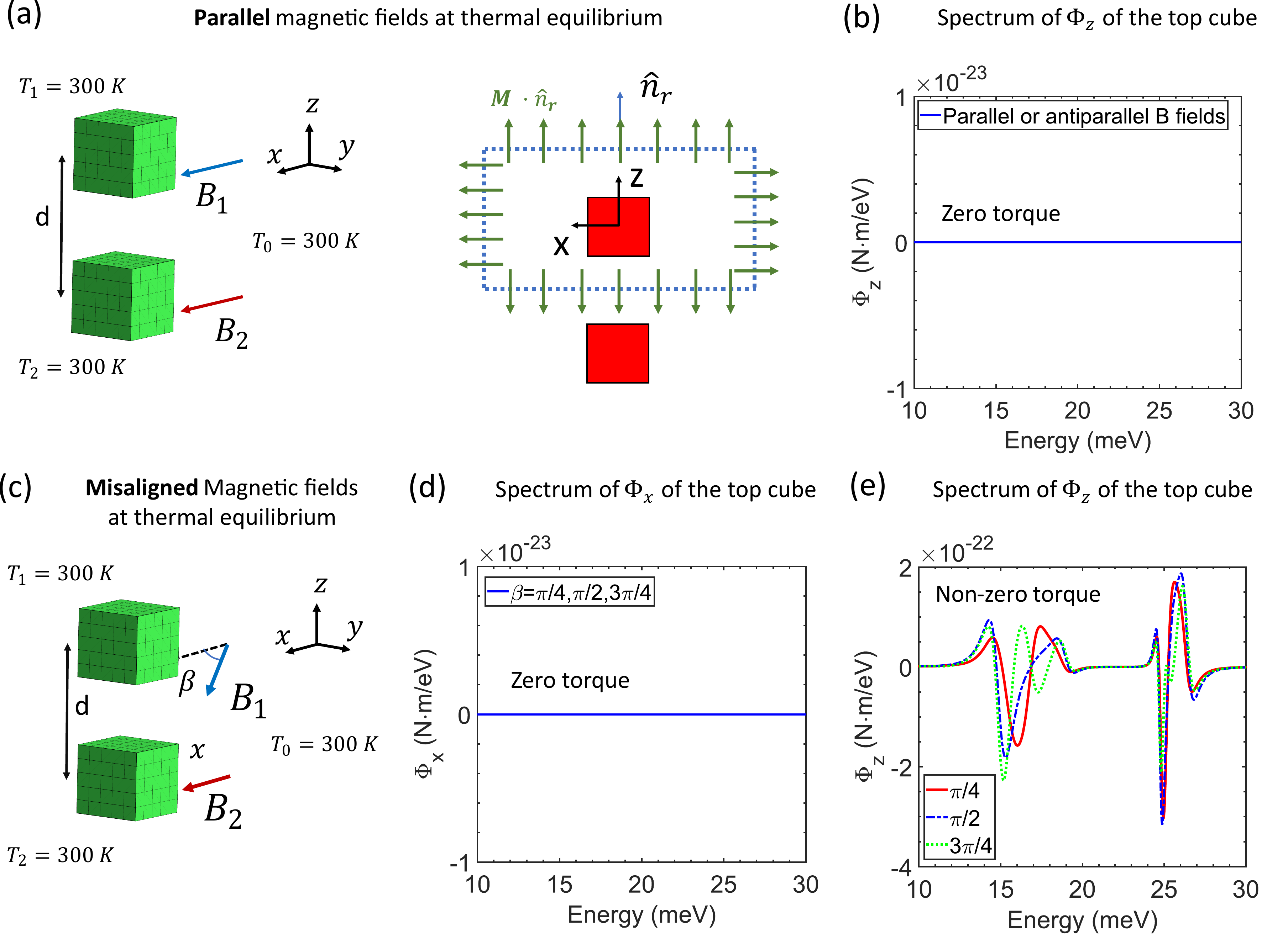}
	\caption{(Color online) Radiative AM transfer in a two-cube system at global thermal equilibrium ($T_1$=$T_2$=$T_0$=300 K). (a) Schematic: The parallel magnetic fields with the same magnitude $B_1$=$B_2$=1 T are applied on both cubes. The size of cubes is L=500 nm and they are separated by d=1 $\mu$m (center to center distance).   (b) The spectrum of the total AM flux $\Phi_z$ emitted out from the top cube. In parallel and anti-parallel magnetic fields, $\Phi_z$ is always zero at global thermal equilibrium. (c) Schematic: The magnetic fields applied on the two cubes have the same magnitude $B_1$=$B_2$=1 T but with an angle $\beta$ between each other. The size of the cubes is L=500 nm and they are separated by d=1 $\mu$m (center distance). (d) and (e) Spectra of the total AM flux $\Phi_i$ (for $i=x,z$, respectively) radiated from the top cube. The spectra are plotted with different magnetic field directions $\beta$. In the misaligned magnetic fields, $\Phi_z$ has a non-zero value while $\Phi_x$ still remains zero. } \label{fig:4}
\end{figure*}

\vspace{2ex}
\textbf{Thermal equilibrium:} We first consider a single cube in equilibrium with vacuum ($T_1=T_0 = 300$ K) and show the balance of the total AM flux $\Phi_{tot}$ in the far field (Fig. \ref{fig:2}). To clarify the origin of the balance, we separate the total flux $\Phi_{tot}$ into two parts: $\Phi_{tot}=\Phi_{p}+\Phi_{E}$. $\Phi_p$ is induced by the fluctuational particle dipoles of the cube that is determined by the body temperature $T_1$ (Appendix \ref{subsec2_2_1:pfl}), while $\Phi_E$ comes from the environmental field fluctuations which are dependent on the environment temperature $T_0$ (Appendix \ref{subsec2_2_2:pf2}). Then we compute each AM flux $\Phi_i$ (for i= x,y,z) across the plane we defined. For this purpose, the AM flux is written as
\begin{equation}
	\Phi_i=\int_{A} \left\langle M_{j i} \right\rangle \mathrm{d} A_{j},
\end{equation}
which describes the integrated flux across a differential section d$A$ perpendicular to the radial vector $\mathbf{R}$. Here we choose a surface that encloses the cube to compute the total AM flux radiated to the far field. As shown in Figure \ref{fig:2}(b), the spectra of the thermal AM flux $\Phi_{x,p}$ and $\Phi_{x,E}$ have the same magnitude but opposite signs at each frequency. It is noted that the background thermal radiation in vacuum has no net flux. The non-zero $\Phi_{x,E}$ originates from the scattering of the incident background thermal radiation by the particle. The total flux $\Phi_{x,tot}$ is zero as $\Phi_{x,p}$ and $\Phi_{x,E}$ perfectly cancel each other. The vanishing of $\Phi_{x,tot}$ satisfies the detailed balance of the AM flux and it indicates that there is no radiative torque applied on the cube at thermal equilibrium. 

For $\Phi_y$ and $\Phi_z$ perpendicular to the gyrotropy axis, both of them are zero regardless of the temperatures and hence result in a zero total flux $\Phi_{i,tot}$ (for $i = y, z$). Since $\Phi_{p}$ and $\Phi_{E}$ have no contribution to the perpendicular component of the AM flux, the torques along y and z directions in this case will always remain zero no matter it is at thermal equilibrium or not. Thus, in the following discussions on the non-equilibrium case of a single cube, we focus on only the AM flux along the magnetic field (gyrotropy axis).

\vspace{2ex}
\textbf{Thermal non-equilibrium:} Here we show that thermal non-equilibrium can lead to a net radiative AM flux along the gyrotropy axis. As shown in Fig. \ref{fig:3} (a), We assume that the environment is at 0 K while the cube is kept at the room temperature $T_1$ = 300 K. Since the temperature of the cube is higher than the environment, $\Phi_p$ has a larger magnitude than $\Phi_E$ at each frequency and thus their summation gives a net AM flux along the magnetic field (Fig. \ref{fig:3}(b)). In addition, due to the zero-point fluctuation, the background radiation still exists at 0 K and thus the AM flux $\Phi_E$ (originating from the environment) in this case has a considerable value. Figure \ref{fig:3}(d) and (e) depict the spatial distributions of the AM flux across yz-planes on the two sides of the cube. AM flux at 24.75 meV is shown here as a typical example. The AM loss through the two planes (in front of and behind the cube) has an equal contribution to the total AM flux, giving a non-zero value of $\Phi_x$. The other components that are perpendicular to the gyrotropy axis ($\Phi_y$ and $\Phi_z$) still remain zero at thermal non-equilibrium. In total, a torque along the external magnetic fields is induced by the AM loss due to the thermal non-equilibrium between the cube and environment.

In Figure \ref{fig:3}(c), we compute the thermal AM torque as a function of the cube size, and compare our numerical calculations to the dipole approximation. In TDDA-125, we divide each cube into 125 subvolumes, while for TDDA-1, we use a single point dipole to represent the object. At small sizes, the results from TDDA-1 and TDDA-125 show a good agreement. When the size of the cube increases, TDDA-125 has a different magnitude as it takes the shape effect into consideration, which should be more precise than TDDA-1 for larger-size objects. Therefore, for following examples, we focus on only the numerical results obtained by TDDA-125.

\subsection{\label{Sec3_2:Two} Thermal AM transfer in a two-cube system}

In the previous section, we showed that, for a single cube, there is no net AM flux going to the far field at thermal equilibrium. The non-vanishing AM flux along the gyroelectric axis exists only when the cube and the environment have different temperatures. However, things can be different for a two-body MO system. In a two-cube system, we find that a net AM flux can also exist in the near field between two cubes when the magnetic fields on each cube are misaligned. Such AM transfer induces an equilibrium torque between the cubes, which can be tuned by changing the angle between the magnetic fields. 

\begin{figure}[b]
	\centering
	\includegraphics[width=0.5\textwidth]{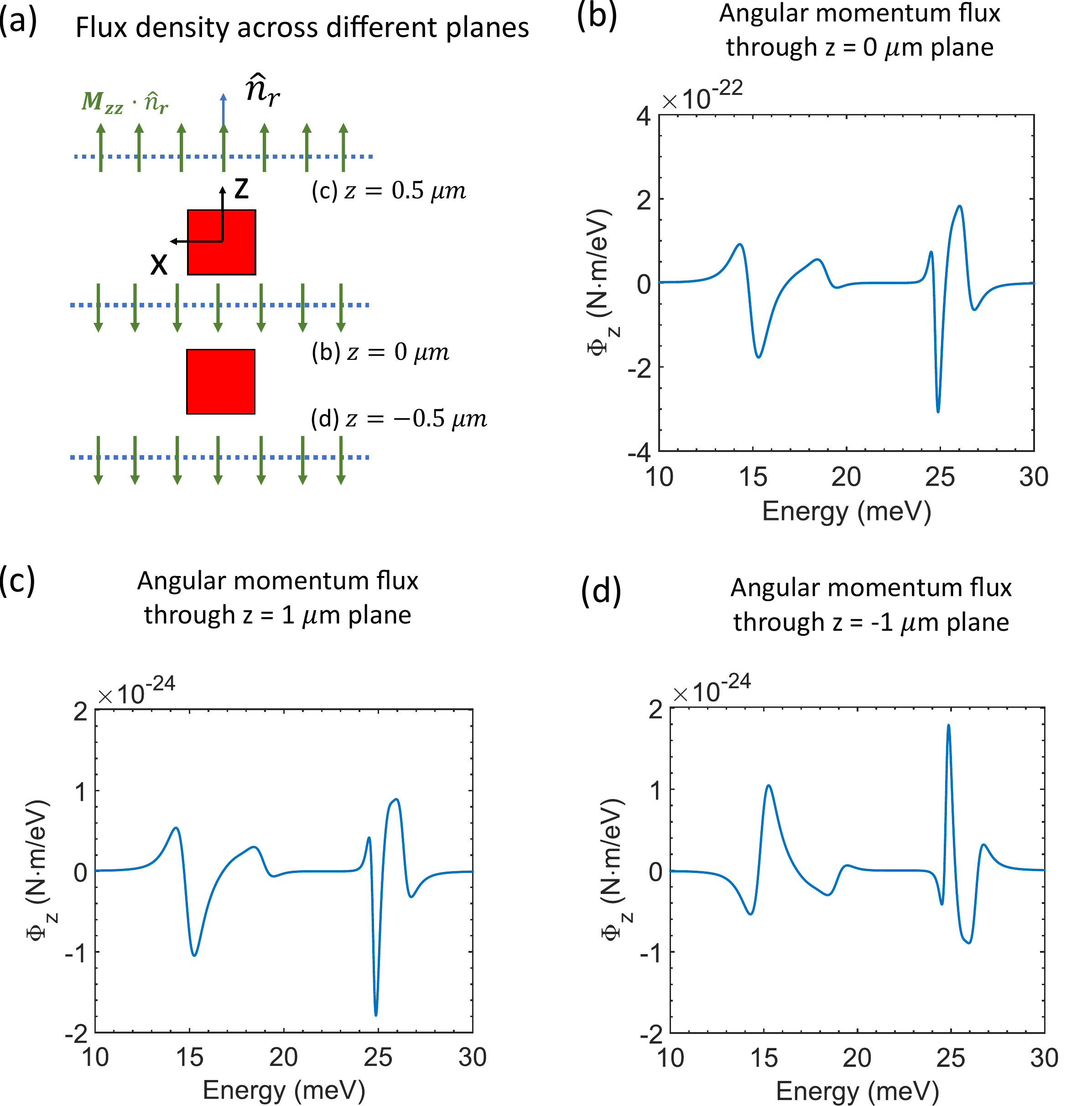}
	\caption{(Color online) Near-field AM flux in a two-cube system at global thermal equilibrium ($T_1$=$T_2$=$T_0$=300 K). The size of cubes is L=500 nm and they are separated by d=1 $\mu$m. The magnitudes of the external magnetic fields are $B_1$=$B_2$=1 T. (a) Schematic. (b)-(d) The spectra of AM flux $\Phi_{z}$ across three different xy-planes as shown in (a). Here $\beta$ is assumed to be $\pi/2$. Figures (c) and (d) show the radiation flux to the environment, and they contribute oppositely to the total flux $\Phi_z$ of the combined two-cube system. At thermal equilibrium, (c) and (d) perfectly cancel each other, giving no net torque on the combined system.} \label{fig:6}
\end{figure}

\vspace{2ex}
\textbf{Thermal equilibrium:} In Fig. \ref{fig:4}, the two-cube system is at global thermal equilibrium where $T_1 = T_2 = T_0 = 300$ K. $T_1$, $T_2$ and $T_0$ denote the temperatures of the top cube, bottom cube and vacuum, respectively. To compute the total AM flux radiating from the top cube, we choose a box enclosing the top cube and calculate the flux across the surface, $\mathbf{M}\cdot\hat{n}_\mathbf{r}$. If the magnetic fields on two cubes are parallel (Fig. \ref{fig:4} (a)-(b)), there is no AM flux going out from either cube, which is similar to the single cube case. However, once the magnetic fields are misaligned, the gyrotropy axes of the cubes are along different directions (Fig \ref{fig:4} (c)-(e)). And then surprisingly, there is a net AM flux along the z direction (perpendicular to the plane formed by the two gyrotropy axes) which is being exchanged between two cubes. Meanwhile, $\Phi_x$ and $\Phi_y$ still remain zero regardless of the angle $\beta$.

\begin{figure*}[t]
	\centering
	\includegraphics[width=1\textwidth]{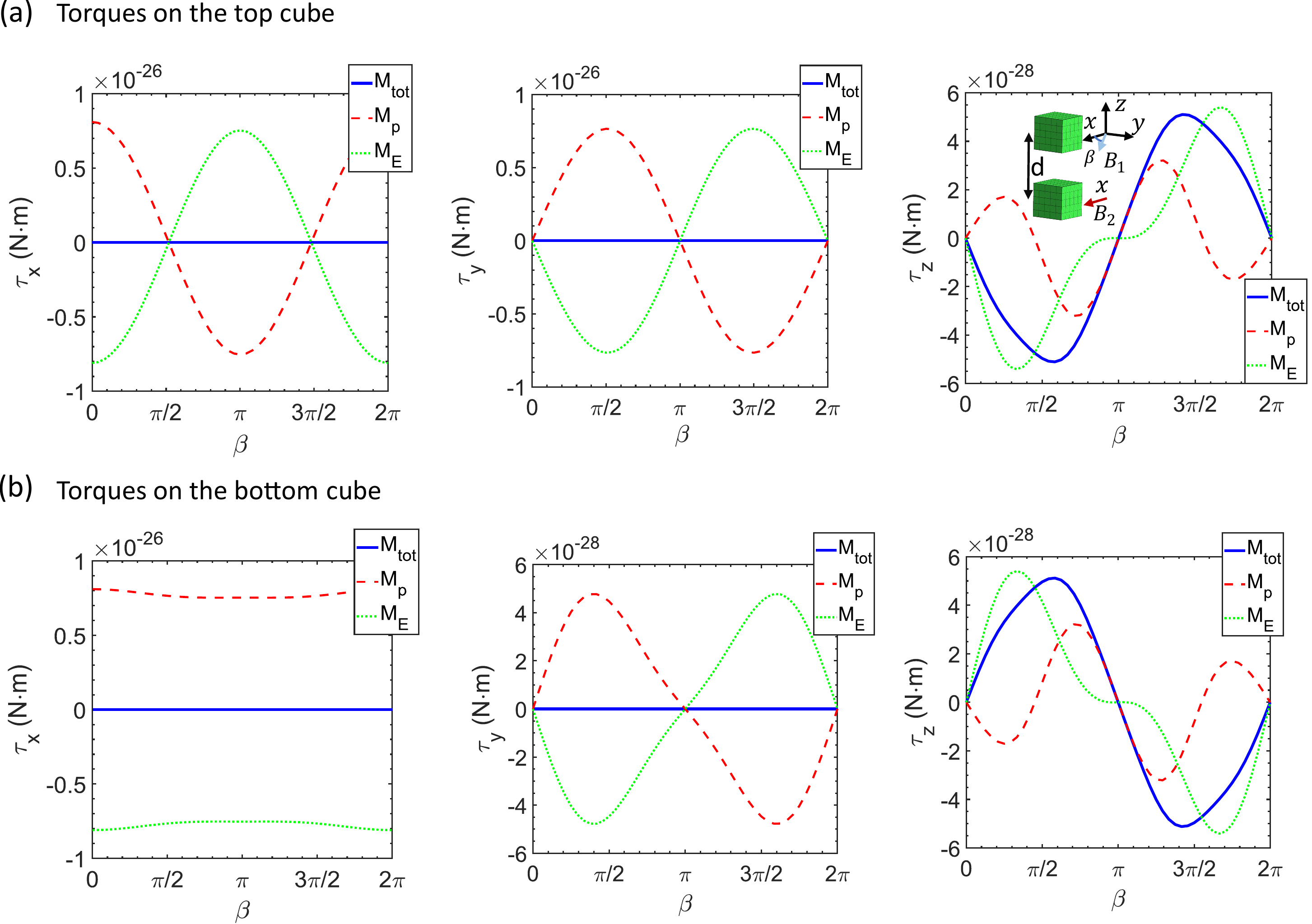}
	\caption{(Color online) Torques induced by radiative AM transfer at global thermal equilibrium ($T_1=T_2=T_0=300$ K) as functions of magnetic field angle $\beta$. The total torque on each cube is separated into two parts according to its origins: $M_p$ is induced by fluctuational dipoles and $M_E$ comes from field fluctuations. $M_{tot}$ is the total torque on each cube that combines $M_p$ and $M_E$. Other parameters are L=500 nm, $B_1$=$B_2$=1 T, and d=1 $\mu$m. (a) Torques on the top cube in a varied magnetic field. (b) Torques on the bottom cube in a fixed magnetic field along x axis.} \label{fig:7}
\end{figure*}

Such AM transfer of $\Phi_z$ is interesting but non-intuitive since it can occur despite the global thermal equilibrium. It is also important to point out that the presence of the non-zero radiative AM transfer at the global thermal equilibrium does not lead to any thermodynamic contradictions. To demonstrate this argument, we separately compute the AM flux $\Phi_z$ across different xy-planes to reveal the origins of the AM transfer (Fig. \ref{fig:6}). Here, we typically choose three planes to show the flux $\Phi_z$: (b) The mid plane between two cubes; (c) the top plane above the top cube; (d) the bottom plane below the bottom cube. The mid xy-plane shows the near field AM transfer between the cubes, while the top and bottom planes show the AM flux going to the far field from each cube, respectively. Here we do not show the AM flux leaking through the other surfaces since their contributions to $\Phi_z$ are negligible compared to xy-planes we showed above. From Fig. \ref{fig:6}, it is easy to find that the AM flux $\Phi_z$ mainly transfers through the mid xy-plane, with almost two-order larger magnitude than the top and bottom planes. Moreover, the fluxes $\Phi_z$ across the top and bottom planes have the same magnitude but different signs (directions) at each frequency. Therefore,considering the total flux radiated to the environment from the combined two-cube system, (c) and (d) cancel each other and give a zero net flux of $\Phi_z$. This means that such AM transfer is a localized phenomenon and won't cause any net flux that transfers between the system and the surrounding environment, conserving AM globally.

With a net AM flux exchanged between the cubes, an equilibrium torque is induced on each cube. Such equilibrium torque can be tuned by changing the angle between the magnetic fields and vanishes when the magnetic fields are parallel. Fig. \ref{fig:7} plots magnetic field dependence of the equilibrium torque. Here we assume that the magnetic field on the bottom cube is fixed while we change the angle $\beta$ of the B field on the top cube. Similar to what we did in the last subsection, we separate the torques into two parts originating from the particle dipole fluctuations ($M_p$) and the environment field fluctuations ($M_E$), respectively.

First we only consider the torques within the xy-plane where the magnetic fields are applied. In this case, $M_p$ and $M_E$ have the same magnitude but different signs and hence result in a zero net torque within the xy-plane on each cube. The vanishing of the torque in xy-planes is independent of $\beta$ and is always true at global thermal equilibrium. Unlike the single-body system where the $M_p$ and $M_E$ should be exactly along its gyrotropy axis (magnetic field), here they can be slightly misaligned with their own gyrotropy axis. Such misalignment originates from the interaction between the cubes since $M_p$ ($M_E$) of the bottom cube in a fixed magnetic field varies when we tune the direction of the B field on the top cube (the left and middle plots in Fig. \ref{fig:7} (b)). The torque that is perpendicular to the external magnetic field has a much smaller magnitude than the parallel component, indicating that the interaction is weak at global thermal equilibrium.

\begin{figure}[t]
	\centering
	\includegraphics[width=0.5\textwidth]{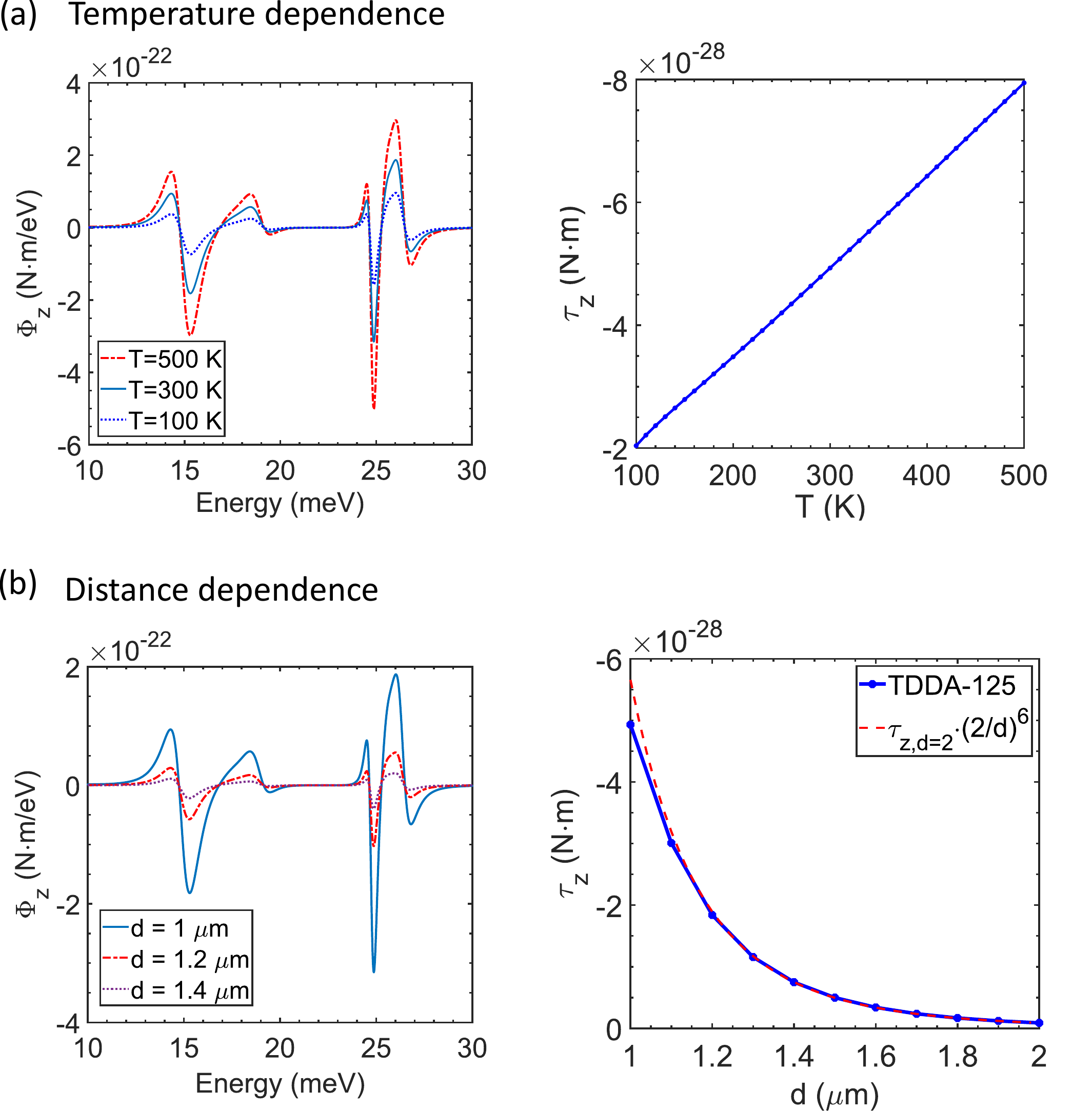}
	\caption{(Color online) (a) Temperature dependence of the near-field AM flux $\Phi_z$ across the mid xy-plane between two cubes. The system remains at global thermal equilibrium $T_1=T_2=T_0$ while tuning the temperature. Here we assume that $\beta=\pi/2$, $d=1$ $\mu$m and $L= 500$ nm. (Left figure) Spectra of $\Phi$ with different temperatures. (Right figure) Induced torque as a function of temperature. (b) Distance dependence of the near-field AM flux $\Phi_z$ that goes through the mid xy-plane between two cubes. The red dash line is plotted by multiplying the torque at d=2 $\mu$m and $(2/d)^6$, which is the ratio to the power of 6 between the torques at two distances.  Parameters are $T_1=T_2=T_0=300$ K, $\beta$=$\pi$/2,  $d=1$ $\mu$m and $L= 500$ nm. (Left figure) Spectra of $\Phi_z$ at different distances. (Right figure) The equilibrium torque $\tau_z$ as a function of distance.} \label{fig:8}
\end{figure}  

Along the z direction perpendicular to the plane formed by the magnetic fields, a torque is induced by the AM transfer between two cubes despite global thermal equilibrium (the right plots in Fig. \ref{fig:7}). When the magnetic fields on two cubes are parallel or anti-parallel, there is no torque applied because the AM transfer is prohibited. Otherwise, each cube feels a torque along the z direction trying to align their gyrotropy axes parallel to each other. Such torques have the same magnitude but opposite signs and, therefore, there is no net torque on the combined system, conserving the global AM at thermal equilibrium. Here we want to note that the "thermal equilibrium" discussed in this work refers to only the global equality of temperature. Due to the nonzero torque induced by the near-field AM transfer, the condition of mechanical equilibrium is broken. The full mechanical dynamics of the particles accounting for all forces including the attractive Casimir force between them is beyond the scope of this work, and will be considered in a future work. 

At thermal equilibrium, there are two distinct zero-torque configurations: parallel gyrotropy axes ($\beta=0$) and anti-parallel gyrotropy axes ($\beta=\pi$). If the system starts from the anti-parallel configuration ($\beta=\pi$), two cubes have a tendency to relax back to the parallel configuration. This means that $\beta=\pi$ is an unstable equilibrium point. On the other hand, if two cubes are left in the parallel configuration ($\beta=0$), they will remain in it and tend to go back after a small disturbance.

Finally, we plot the temperature dependence and distance dependence of the torque $\tau_z$ in Figure \ref{fig:8}. Temperature affects the AM flux through the mean thermal energy $\Theta(\omega, T)=\hbar \omega / 2+\hbar \omega /\left[\exp \left(\hbar \omega / k_{B} T\right)-1\right]$. Since the mean thermal energy is approximately constant over the frequency range of interest, the magnitude of $\tau_z$ increases proportionately with the temperature. The left panel in Fig. \ref{fig:8}(b) demonstrates the spectrum for $\Phi_z$ as a function of distance between two cubes. At each frequency, the sign (the direction of the radiative AM flux) stays the same while the magnitude decays as a function of distance. The right panel in Fig. \ref{fig:8}(b) shows the torque along the z direction after doing integral of $\Phi_z$ over frequency. Similarly, the magnitude of the torque $\tau_z$ decreases as a function of distance while its direction remains unchanged. This confirms that the radiative AM transfer for $\Phi_z$ at global thermal equilibrium originates from the near-field interaction between the cubes, as the interaction strength decays with increasing separation between them.

\vspace{2ex}
\textbf{Thermal non-equilibrium:}  Above we showed that, for a two-cube system at global thermal equilibrium ($T_1=T_2=T_0$), AM transfer can happen only along z direction. And we have also pointed out that there is no torque within xy-plane at global thermal equilibrium due to the balance between $M_p$ and $M_E$, which are slightly misaligned with their gyrotropy axes. Here we are going to show that non-equilibrium ($T_1\neq T2$) helps amplify the interaction between the cubes, making the AM transfer between two cubes much stronger than the equilibrium case. 

\begin{figure}[t]
	\centering
	\includegraphics[width=0.5\textwidth]{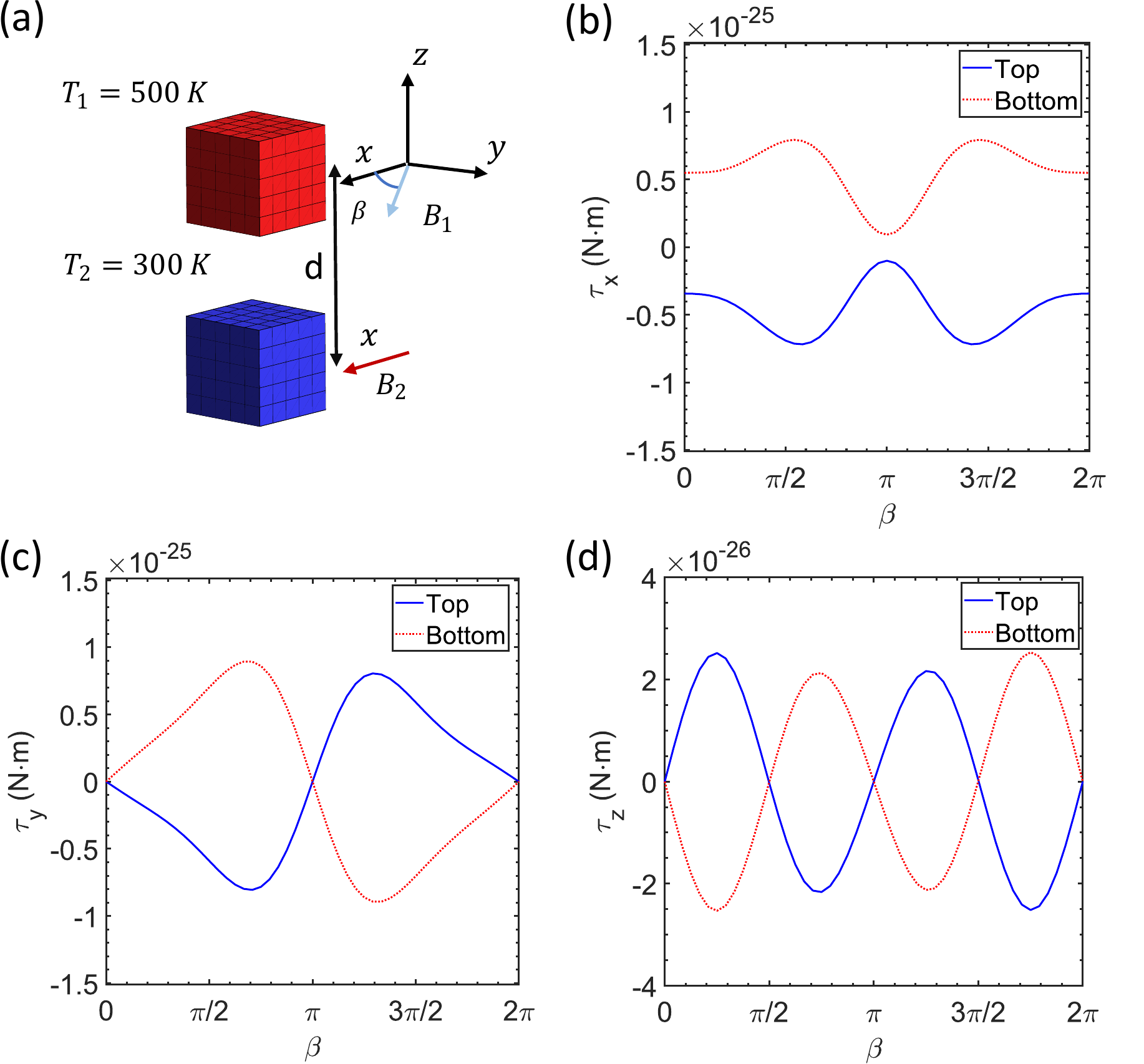}
	\caption{(Color online) Thermal non-equilibrium torques when the top cube is heated up to $T_1$=500 K and the bottom cube still remains at thermal equilibrium with the surroundings, $T_2=T_0=300$ K. The size of the cubes is $L= 500$ nm and they are separated by $d=1$ $\mu$m. (a) Schematic of the thermal non-equilibrium case. (b)-(d) Torques as functions of the direction of the magnetic field. The magnetic field on the bottom cube is fixed, while the direction of the magnetic field on the top cube is varied from 0 to 2$\pi$.} \label{fig:9}
\end{figure}

Figure \ref{fig:9} depicts the magnetic field dependence of the radiative AM torques at thermal non-equilibrium. As shown in Figure \ref{fig:9} (a), we assume that the top cube is heated up to $T_1=500$ K, while the bottom cube remains	at equilibrium with the environment ($T_2=T_0=300$ K). Similarly, we tune the direction of the magnetic field on the top cube and compute the non-equilibrium torque as a function of $\beta$. In this case, the torques on each cube are strongly modified due to the near-field AM transfer (Figure \ref{fig:9} (b)-(d)). First, the torques within the xy-plane no longer follow the direction of the external magnetic fields. On the contrary, the torques in the xy-pane have a big departure from the gyrotropy axes. Second, the torque in the xy-plane on the bottom cube (in a fixed magnetic field) also shows a strong dependence of the magnetic field $B_1$ applied on the top cube. At each angle $\beta$, the torques on two cubes have comparable magnitudes and opposite signs. Their summation gives a smaller total torque on the combined system, which is one order smaller than the torque on each cube. Third, the torques along the z direction are amplified by two orders compared with the equilibrium case, directly indicating that the radiative AM transfer is strongly amplified because of the unequal temperatures of two cubes.

\section{Additional remarks AND Conclusions } \label{Sec:Conclusion}

We demonstrated radiative AM transfer different from typically considered heat (energy) transfer in nonreciprocal systems. We have developed the TDDA approach based on fluctuational electrodynamics for analyzing the thermal AM flux density in the near-field and far-field of finite non-reciprocal bodies. Our work reveals that the AM loss due to far-field radiation plays a fundamental role in generating a thermal AM torque along the gyroelectric axis of a single non-reciprocal body at thermal non-equilibrium. The connection between the thermal radiation of nonreciprocal bodies and the induced torques is important for exploring new ways of directional thermal AM transfer. 

We also found that in a nonreciprocal system, the near-field thermal AM between two objects can be nonzero despite the zero heat (energy) transfer.  This is a localized phenomenon that happens because of near-field interaction between the two bodies and decays with increasing distance between the bodies. Also, such near-field interaction will not contribute to any net radiative AM flux at far field for the combined system. Moreover, the near-field AM flux across the plane between two bodies is not necessary to be always along their gyrotropy axes. With misaligned gyroelectric axes, an AM perpendicular to the plane formed by two gyrotropy axes can transfer between two bodies despite global thermal equilibrium. It induces torques on both cubes trying to align two gyrotropy axes parallel to each other. At global thermal equilibrium, the torques on two cubes have the same magnitude but opposite signs, and the net torque on the combined system is zero, satisfying the detailed balance of AM exchange between the combined system and the environment at equilibrium.  

Some recent works predicted that the torques can be induced by vacuum friction of a rotating object made of reciprocal isotropic media \cite{manjavacas2010thermal,zhao2012rotational,manjavacas2010vacuum,zhujing2020enhancement}. These predictions are of high interest in the context of nanophotonics \cite{shitrit2013spin,wu2014spectrally,khandekar2019circularly,ahn2020ultrasensitive,guo2020single,hoang2016torsional,ahn2018optically} and Casimir physics \cite{somers2018measurement,xu2017detecting,guerout2015casimir}, as they fundamentally originate from quantum and thermal fluctuations. However, such rotational vacuum frictions still remain at the theoretical level and, so far, no experimental observation has been done due to the extremely small magnitudes of the torques which are far below the sensitivity that can be achieved by the torque sensor \cite{kim2016approaching}. Many theoretical works tried to enhance the vacuum friction to make it more measurable than a single rotating particle \cite{manjavacas2010thermal}, such as using surface plasmon resonance \cite{zhao2012rotational} and surface photon tunneling \cite{zhujing2020enhancement}. Recently, an experimental demonstration of the most sensitive torque measurement was done with a levitated nanoparticle, improving the torque sensitivity by a few orders \cite{ahn2020ultrasensitive} and showing the feasibility of detecting the rotational vacuum friction near a surface. And as we have shown in this work, the thermal AM torque in a MO system in the presence of external magnetic fields has a magnitude comparable to with the sensitivity that can be achieved by the levitated torque sensor \cite{ahn2020ultrasensitive} and it is tunable by changing the external magnetic fields. Therefore, the thermal AM torques in the MO system will be experimentally detectable in the near future.

Finally, the TDDA approach we have developed in this work can be applied to describe the radiative AM transfer of finite objects with arbitrary size and shape. This formalism allows us to compute the radiative AM flux in the near field as well as far field of the MO bodies with arbitrary permittivity tensors. We have used this TDDA approach to explore how the nonreciprocity affects the radiation of AM in single-cube and two-cube systems. Our work provides a way for to describe AM-resolved thermal radiation involving finite MO bodies of arbitrary shape.

\begin{acknowledgments}
We are grateful for the support from the DARPA QUEST program and the Office of Naval Research under Grant No. N00014-18-1-2371.
\end{acknowledgments}

\appendix

\section{TDDA Approach to Radiative AM Flux} \label{Append:TDDA}
In this section, we develop a TDDA method for calculating the AM flux density based on former works \cite{edalatpour2014thermal,edalatpour2015convergence,ekeroth2017thermal}. Here we separate the discussion into two parts: particle dipole fluctuations and environmental field fluctuations. These two contributions correspond to the emission of objects and the absorption from the surrounding environment, respectively. And they will be balanced when the system is in equilibrium with the environment. 

In particular, here we consider the case of two finite objects assumed to be at fixed temperatures $T_1$ and $T_2$ and they both interact with a thermal bath at temperature $T_0$. We assume that these two objects are described by a collection of $N_p$ (for object p) electric point dipoles. Each dipole is characterized by a volume $V_{i,p}$ and a polarizability tensor $\hat{\alpha_{i,p}}$, where p=1,2 denotes the body that the dipole belongs to and i=1,2,...,$N_p$ indicates the i-th subvolume in that object. We group the electric dipoles and electric fields in a compact form
\begin{equation}
\begin{array}{l}
\overline{\mathbf{P}}=\left(\begin{array}{c}
\overline{\mathbf{P}}_{1} \\
\overline{\mathbf{P}}_{2}
\end{array}\right) ; \quad \overline{\mathbf{P}}_{1}=\left(\begin{array}{c}
\mathbf{p}_{1,1} \\
\vdots \\
\mathbf{p}_{N_{1}, 1}
\end{array}\right), \overline{\mathbf{P}}_{2}=\left(\begin{array}{c}
\mathbf{p}_{1,2} \\
\vdots \\
\mathbf{p}_{N_{2}, 2}
\end{array}\right) \\
\overline{\mathbf{E}}=\left(\begin{array}{c}
\overline{\mathbf{E}}_{1} \\
\overline{\mathbf{E}}_{2}
\end{array}\right) ; \quad \overline{\mathbf{E}}_{1}=\left(\begin{array}{c}
\mathbf{E}_{1,1} \\
\vdots \\
\mathbf{E}_{N_{1}, 1}
\end{array}\right), \overline{\mathbf{E}}_{2}=\left(\begin{array}{c}
\mathbf{E}_{1,2} \\
\vdots \\
\mathbf{E}_{N_{2}, 2}
\end{array}\right)
\end{array}
\end{equation}
Then we can define the polarizability tensor as $\bar{\alpha}=\begin{pmatrix}
\bar{\alpha}_1& 0\\ 
0&\bar{\alpha}_2 
\end{pmatrix}$ and $\bar{\alpha}_p=$ diag$(\hat{\alpha}_{1, p}, \ldots, \hat{\alpha}_{N_{p}, p})$, ($p$ = 1,2). And each element $\hat{\alpha}_{i, p}$ is given by \cite{ekeroth2017thermal}
\begin{equation}
\hat{\alpha}_{i,p}=\left(\frac{1}{V_p}\left(\hat{L}_{p}+\left[\hat{\epsilon}_{p}-\mathbb{I}\right]^{-1}\right)-i \frac{k_{0}^{3}}{6 \pi} \mathbb{I}\right)^{-1}
\end{equation}
where $\hat{\epsilon}_{p}$ is the dielectric permittivity tensor, $V_p$ is the volume of each discrete dipole, and $\hat{L_p}$ is the depolarization tensor which is $\hat{L_p}=(1/3)\mathbb{I}$ for the cubic volume mesh element \cite{ekeroth2017thermal,lakhtakia1992general,yaghjian1980electric}.

To compute the total radiative momentum and AM flux that are described in terms of the surface integral of the flux densities as Eq. (\ref{Eq:force}) and Eq. (\ref{Eq:torque}), we need to compute the statistical average of the momentum and AM flux density by Eq. (\ref{Eq:MaxwellStress}) and Eq. (\ref{Eq:AngularDensity}). For ease of analysis, the average is expressed in terms of the Fourier transforms
\begin{equation} \label{Eq:TDDA_Sigma}
\begin{aligned}
\left\langle \mathbf{\Sigma}(\mathbf{r})\right\rangle=& -2 \int_{0}^{\infty} \frac{d \omega}{2 \pi} \int_{-\infty}^{\infty} \frac{d \omega^{\prime}}{2 \pi}\\
&\operatorname{Re}[ \epsilon_0\left\langle\mathbf{E}(\mathbf{r}, \omega) \otimes \mathbf{E}^*(\mathbf{r}, \omega^{\prime})\right\rangle e^{-i\left(\omega-\omega^{\prime}\right) t}\\
&+\mu_0\left\langle\mathbf{H}(\mathbf{r}, \omega) \otimes \mathbf{H}^*(\mathbf{r}, \omega^{\prime})\right\rangle e^{-i\left(\omega-\omega^{\prime}\right) t}\\
&-\frac{1}{2}\epsilon_0\operatorname{Tr}\left\{\left\langle\mathbf{E}(\mathbf{r}, \omega) \otimes \mathbf{E}^*(\mathbf{r}, \omega^{\prime})\right\rangle e^{-i\left(\omega-\omega^{\prime}\right) t}\right\}\mathbb{I}\\
&-\frac{1}{2}\mu_0\operatorname{Tr}\left\{\left\langle\mathbf{H}(\mathbf{r}, \omega) \otimes \mathbf{H}^*(\mathbf{r}, \omega^{\prime})\right\rangle e^{-i\left(\omega-\omega^{\prime}\right) t}\right\}\mathbb{I}]
\end{aligned}
\end{equation}
for momentum flux density and
\begin{equation} \label{Eq:TDDA_tau}
\begin{aligned}
\left\langle \mathbf{M}(\mathbf{r})\right\rangle=& -2 \int_{0}^{\infty} \frac{d \omega}{2 \pi} \int_{-\infty}^{\infty} \frac{d \omega^{\prime}}{2 \pi}\mathbf{r}\times\\
&\operatorname{Re}[ \epsilon_0\left\langle\mathbf{E}(\mathbf{r}, \omega) \otimes \mathbf{E}^*(\mathbf{r}, \omega^{\prime})\right\rangle e^{-i\left(\omega-\omega^{\prime}\right) t}\\
&+\mu_0\left\langle\mathbf{H}(\mathbf{r}, \omega) \otimes \mathbf{H}^*(\mathbf{r}, \omega^{\prime})\right\rangle e^{-i\left(\omega-\omega^{\prime}\right) t}\\
&-\frac{1}{2}\epsilon_0\operatorname{Tr}\left\{\left\langle\mathbf{E}(\mathbf{r}, \omega) \otimes \mathbf{E}^*(\mathbf{r}, \omega^{\prime})\right\rangle e^{-i\left(\omega-\omega^{\prime}\right) t}\right\}\mathbb{I}\\
&-\frac{1}{2}\mu_0\operatorname{Tr}\left\{\left\langle\mathbf{H}(\mathbf{r}, \omega) \otimes \mathbf{H}^*(\mathbf{r}, \omega^{\prime})\right\rangle e^{-i\left(\omega-\omega^{\prime}\right) t}\right\}\mathbb{I}]
\end{aligned}
\end{equation}
for AM flux density. Here we have used the properties that $\mathbf{E}(\mathbf{r}, -\omega)=\mathbf{E}(\mathbf{r}, \omega)^*$ and $\mathbf{H}(\mathbf{r}, -\omega)=\mathbf{H}(\mathbf{r}, \omega)^*$. Using FDT from Eq. \ref{Eq:FDT1} and Eq. \ref{Eq:FDT2}, the above expression can be reduced to the integration of the terms containing $\left\langle\mathbf{E}(\mathbf{r}, \omega) \otimes \mathbf{E}^*(\mathbf{r}, \omega)\right\rangle$ and $\left\langle\mathbf{H}(\mathbf{r}, \omega) \otimes \mathbf{H}^*(\mathbf{r}, \omega)\right\rangle$. In most cases in the remaining discussions, we drop the argument $\omega$ to alleviate the notation. In the next two subsections, we will separately formalize the TDDA approach for computing the AM flux due to the electric dipole fluctuations and the electromagnetic field fluctuations.

\subsection{\label{subsec2_2_1:pfl} Electric Dipole Fluctuation}
We start by decomposing the local field $\mathbf{E}(\mathbf{r})$ outside the objects into the source field $\mathbf{E}_0(\mathbf{r})$ and the induced part $\mathbf{E}^{ind}(\mathbf{r})$:
\begin{equation}\label{Eq:Efield_decom}
\mathbf{E}(\mathbf{r})=\mathbf{E}_0(\mathbf{r})+\mathbf{E}^{ind}(\mathbf{r}).
\end{equation}
The source field $\mathbf{E}_0(\mathbf{r})$ originates from the fluctuating dipole inside the objects and is given by
\begin{equation}
\mathbf{E}_{0}(\mathbf{r})=\frac{k_{0}^{2}}{\epsilon_{0}} \mathbf{G}_{EE} \bar{\mathbf{P}}_{\mathrm{f}},
\end{equation}
where $\mathbf{G}_{EE}=\left(\hat{G}_{EE}\left(\mathbf{r}, \mathbf{r}_{1}\right), \ldots, \hat{G}_{EE}\left(\mathbf{r}, \mathbf{r}_{N}\right)\right)$ is the row vector of the free-space Green tensors:
\begin{equation}\begin{aligned} \label{Eq:DyadicGreens}
\hat{G}_{EE}\left(\mathbf{r}, \mathbf{r}^{\prime}\right)=& \frac{e^{i k_{0} R}}{4 \pi R}\left[\left(1+\frac{i k_{0} R-1}{k_{0}^{2} R^{2}}\right) \hat{1}\right.\\
 &+\left.\left(\frac{3-3 i k_{0} R-k_{0}^{2} R^{2}}{k_{0}^{2} R^{2}}\right) \frac{\mathbf{R} \otimes \mathbf{R}}{R^{2}}\right],
\end{aligned}\end{equation}
where $\mathbf{R}=\mathbf{r}-\mathbf{r}^{\prime}$ and $R=\left|\mathbf{r}-\mathbf{r}^{\prime}\right|$. $\otimes$ denotes the outer product of two vectors. The dyadic Green's tensor connects the fluctuating dipoles to the observation point outside the objects. $\bar{\mathbf{P}}_{\mathrm{f}}$ is the fluctuating dipoles inside the objects and can be obtained by FDT
\begin{equation}
\left\langle\bar{\mathbf{P}}_{\mathrm{f}}(\omega) \bar{\mathbf{P}}_{\mathrm{f}}^{\dagger}\left(\omega^{\prime}\right)\right\rangle=2 \pi\hbar \epsilon_{0}  \delta\left(\omega-\omega^{\prime}\right)\left[\mathbb{I}+2 \hat{n}_{\mathrm{B}}\left(\omega, T_1, T_2\right)\right] \hat{\chi}
\end{equation}
where $ \hat{n}_{\mathrm{B}}\left(\omega, T_1, T_2\right)$ is a diagonal tensor with $3N$ elements given by the Bose-Einstein distribution
\begin{equation}
\hat{n}_{\mathrm{B}}\left(\omega, T_1, T_2\right)=\begin{pmatrix}
n_1(\omega)\mathbb{I}_{3N_1\times 3N_1}&\mathbf{\hat{0}}\\ 	
\mathbf{\hat{0}} & n_2(\omega)\mathbb{I}_{3N_2\times 3N_2}
\end{pmatrix}.
\end{equation}
and we have also introduced 
\begin{equation}
\hat{\chi}=\frac{1}{2 i}\left(\hat{\alpha}-\hat{\alpha}^{\dagger}\right)-\frac{k_{0}^{3}}{6 \pi} \hat{\alpha}^{\dagger} \hat{\alpha}.
\end{equation}
as the radiative correction \cite{manjavacas2012radiative,messina2013fluctuation}. 

The second term in Eq. (\ref{Eq:Efield_decom}) originates from the induced dipoles: $\mathbf{E}^{ind}(\mathbf{r})=\frac{k_{0}^{2}}{\epsilon_{0}} \mathbf{G}_{EE} \bar{\mathbf{P}}^{ind}$. And the induced dipole $\bar{\mathbf{P}}^{ind}$ comes from the electric field inside the bodies as 
$\bar{\mathbf{P}}^{ind}=\epsilon_0 \bar{\alpha}\bar{\mathbf{E}}$, while $\bar{\mathbf{E}}$ can be obtained by solving TDDA equation
\begin{equation}\label{Eq:TDDA}
\bar{\mathbf{E}}=\bar{\mathbf{E}}_0+k_0^{2} \textup{d}\bar{\mathbb{G}}_{EE}\bar{\alpha}\bar{\mathbf{E}}.
\end{equation}
where $\textup{d}\bar{\mathbb{G}}_{EE}=\bar{\mathbb{G}}_{EE}-\operatorname{diag}\{\bar{\mathbb{G}}_{EE}\}$ and $\bar{\mathbb{G}}_{EE}$ is the matrix of the dynadic Green's tensors inside bodies that is also defined by Eq. (\ref{Eq:DyadicGreens}). Here the overhead bar notation $"\ \bar{\ }\ "$ indicates that the quantities are evaluated inside the bodies and it will be the same for other quantities in the following discussion.
Solving Eq. (\ref{Eq:TDDA}), we get the induced electric field
\begin{equation}
\mathbf{E}^{ind}(\mathbf{r})=\frac{k_{0}^{2}}{\epsilon_{0}} \mathbf{G}_{EE} \bar{\mathbf{T}}^{-1} \bar{\mathbf{P}}_{\mathrm{f}}
\end{equation}
where 
\begin{equation} \label{Eq:E_p_TDDA}
\mathbf{T}_{i j}=\delta_{i j} \mathbb{I}-\left(1-\delta_{i j}\right) k_0^{2} \bar{\alpha_{i}} \bar{\mathbb{G}}_{EE,i j}
\end{equation}

Now, making use of Eq. (\ref{Eq:E_p_TDDA}), we can compute the electric field correlation $\left\langle\mathbf{E}(\mathbf{r}, \omega) \otimes \mathbf{E}^*(\mathbf{r}, \omega)\right\rangle$  which is required for obtaining AM flux:
\begin{equation} \label{Eq:E_cor_p}
\left\langle\mathbf{E}(\mathbf{r}, \omega) \otimes \mathbf{E}^*(\mathbf{r}, \omega)\right\rangle=\frac{k_0^4}{\epsilon_{0}^2}\mathbf{G}_{EE}\bar{\mathbf{T}}^{-1}\left\langle\bar{\mathbf{P}}_{\mathrm{f}} \bar{\mathbf{P}}_{\mathrm{f}}^{\dagger}\right\rangle\bar{\mathbf{T}}^{-1\dagger}\mathbf{G}_{EE}^{\dagger}
\end{equation}
Similarly, the magnetic field correlation $\left\langle\mathbf{H}(\mathbf{r}, \omega) \otimes \mathbf{H}^*(\mathbf{r}, \omega)\right\rangle$ can be obtained by replacing the electric-electric dyadic Green's tensor by the magnetic-electric Green's tensor:
\begin{equation}\label{G_HE}
\begin{aligned}
\hat{G}_{HE}\left(\mathbf{r}, \mathbf{r}^\prime\right)=\frac{e^{i k_0 R}}{4\pi R}\left(1+\frac{i}{k_0 r}\right)\sqrt{\frac{\epsilon_0}{\mu_0}}\left[\begin{array}{ccc}
0 & -\hat{\mathbf{r}}_{z} & \hat{\mathbf{r}}_{y} \\
\hat{\mathbf{r}}_{z} & 0 & -\hat{\mathbf{r}}_{x} \\
-\hat{\mathbf{r}}_{y} & \hat{\mathbf{r}}_{x} & 0
\end{array}\right]
\end{aligned}
\end{equation}
where $\hat{\mathbf{r}}=\mathbf{R}/R$, ($i$ = $x$,$y$,$z$). Then we have
\begin{equation}  \label{Eq:H_cor_p}
\left\langle\mathbf{H}(\mathbf{r}, \omega) \otimes \mathbf{H}^*(\mathbf{r}, \omega)\right\rangle=\frac{k_0^4}{\epsilon_{0}^2}\mathbf{G}_{HE}\bar{\mathbf{T}}^{-1}\left\langle\bar{\mathbf{P}}_{\mathrm{f}} \bar{\mathbf{P}}_{\mathrm{f}}^{\dagger}\right\rangle\bar{\mathbf{T}}^{-1\dagger}\mathbf{G}_{HE}^{\dagger}
\end{equation}

\subsection{\label{subsec2_2_2:pf2} Electromagnetic Field Fluctuation}

Considering the interaction with the thermal bath, the fluctuating electromagnetic field will lead to the AM transfer and, to some extent, balances the contribution from the electric dipole fluctuation. The fluctuating electromagnetic field also fulfills the FDT which gives:
\begin{equation}
\left\langle\bar{\mathbf{E}}_{\mathrm{f}}(\omega) \bar{\mathbf{E}}_{\mathrm{f}}^{\dagger}\left(\omega^{\prime}\right)\right\rangle=2 \pi \frac{\hbar k_0^2}{\epsilon_{0} } \delta\left(\omega-\omega^{\prime}\right)\left(1+2 n_{\mathrm{0}}\left(\omega\right)\right) \operatorname{Im}\bar{\mathbb{G}}_{EE}
\end{equation}

\begin{equation} \label{Eq:correlat1}
\left\langle\bar{\mathbf{E}}_{\mathrm{f}}(\omega) \mathbf{E}_{\mathrm{f}}^{\dagger}\left(\omega^{\prime}\right)\right\rangle=2 \pi\frac{\hbar k_0^2}{\epsilon_{0} }  \delta\left(\omega-\omega^{\prime}\right)\left(1+2 n_{\mathrm{0}}\left(\omega\right)\right) \operatorname{Im}\mathbf{G}_{EE}
\end{equation}

\begin{equation} \label{Eq:correlat2}
\left\langle\mathbf{H}_{\mathrm{f}}(\omega) \bar{\mathbf{E}}_{\mathrm{f}}^{\dagger}\left(\omega^{\prime}\right)\right\rangle=2 \pi\frac{\hbar k_0^2}{\epsilon_{0} }  \delta\left(\omega-\omega^{\prime}\right)\left(1+2 n_{\mathrm{0}}\left(\omega\right)\right) \operatorname{Im}\mathbf{G}_{HE}
\end{equation}
It is noted that here we also considered the correlation between fluctuating fields inside bodies and at observation point outside bodies by Eq. (\ref{Eq:correlat1}) and Eq. (\ref{Eq:correlat2}). That is because the induced electric dipoles generate electric/magnetic field at the observation point that correlates to the local fluctuating field. However, we have ignored the correlation between fluctuating magnetic fields due to the unit permeability ($\mu=1$) of the media we are studying.

As usual, we decompose the local electric field at observation point into the source field and the induced field 
\begin{equation}\label{Eq:Efield_decom2}
\mathbf{E}(\mathbf{r})=\mathbf{E}_0(\mathbf{r})+\mathbf{E}^{ind}(\mathbf{r}).
\end{equation}
And the source field $\mathbf{E}_0$ in this case is the bosonic field of the thermal bath so that $\mathbf{E}_0=\mathbf{E}_f$. And the second term is the induced electric field from the induced dipoles $\mathbf{E}^{ind}(\mathbf{r})=\frac{k_{0}^{2}}{\epsilon_{0}} \mathbf{G}_{EE} \bar{\mathbf{P}}^{ind}$. The induced dipole $\bar{\mathbf{P}}^{ind}$ in this case can be obtained by solving the TDDA equation for electric dipole
\begin{equation}
\bar{\mathbf{P}}^{ind}=\bar{\mathbf{P}}_0+k_0^2\bar{\alpha}\textup{d}\bar{\mathbb{G}}_{EE}\bar{\mathbf{P}}^{ind}
\end{equation}
where $\bar{\mathbf{P}}_0=\epsilon_0 \bar{\alpha} \bar{\mathbf{E_f}}$ is directly induced by the local fluctuating field. Then we write the total electric field $\mathbf{E}(\mathbf{r})$ as
\begin{equation}\label{Eq:E_E_tot}
\mathbf{E}=\mathbf{E}_f+k_0^2\bar{\alpha}\mathbf{G}_{EE}T^{-1}\bar{\alpha}\bar{\mathbf{E}}_f.
\end{equation}
The first term $\mathbf{E}_f$ on the left hand side (LHS) denotes the local fluctuating field at the observation point. $\bar{\mathbf{E}}_f$ in the second term of LHS is the fluctuating field inside bodies. The correlation of electric field at observation point is obtained with the help of Eq. (\ref{Eq:E_E_tot}):
\begin{equation} \label{Eq:E_cor_E}
\begin{aligned}
\left\langle\mathbf{E}(\mathbf{r}, \omega) \otimes \mathbf{E}^*(\mathbf{r}, \omega)\right\rangle=&k_0^4\mathbf{G}_{EE}\bar{\mathbf{T}}^{-1}\bar{\alpha}\left\langle\bar{\mathbf{E}}_{\mathrm{f}} \bar{\mathbf{E}}_{\mathrm{f}}^{\dagger}\right\rangle\bar{\alpha}^\dagger\bar{\mathbf{T}}^{-1\dagger}\mathbf{G}_{EE}^{\dagger}\\
&+k_0^2\mathbf{G}_{EE}\bar{\mathbf{T}}^{-1}\bar{\alpha}\left\langle\bar{\mathbf{E}}_{\mathrm{f}} \mathbf{E}_{\mathrm{f}}^{\dagger}\right\rangle\\
&+k_0^2\left\langle\mathbf{E}_{\mathrm{f}} \bar{\mathbf{E}}_{\mathrm{f}}^{\dagger}\right\rangle\bar{\alpha}^\dagger\bar{\mathbf{T}}^{-1\dagger}\mathbf{G}_{EE}^{\dagger}
\end{aligned}
\end{equation}

Similarly, replacing $\mathbf{G}_{EE}$ with $\mathbf{G}_{HE}$, we can get the correlation of magnetic field
\begin{equation} \label{Eq:H_cor_E}
\begin{aligned}
\left\langle\mathbf{H}(\mathbf{r}, \omega) \otimes \mathbf{H}^*(\mathbf{r}, \omega)\right\rangle=&k_0^4\mathbf{G}_{HE}\bar{\mathbf{T}}^{-1}\bar{\alpha}\left\langle\bar{\mathbf{E}}_{\mathrm{f}} \bar{\mathbf{E}}_{\mathrm{f}}^{\dagger}\right\rangle\bar{\alpha}^\dagger\bar{\mathbf{T}}^{-1\dagger}\mathbf{G}_{HE}^{\dagger}\\
&+k_0^2\mathbf{G}_{HE}\bar{\mathbf{T}}^{-1}\bar{\alpha}\left\langle\bar{\mathbf{E}}_{\mathrm{f}} \mathbf{H}_{\mathrm{f}}^{\dagger}\right\rangle\\
&+k_0^2\left\langle\mathbf{H}_{\mathrm{f}} \bar{\mathbf{E}}_{\mathrm{f}}^{\dagger}\right\rangle\bar{\alpha}^\dagger\bar{\mathbf{T}}^{-1\dagger}\mathbf{G}_{HE}^{\dagger}
\end{aligned}
\end{equation}

Now, combining Eq. (\ref{Eq:TDDA_tau}), (\ref{Eq:E_cor_p}), (\ref{Eq:H_cor_p}), (\ref{Eq:E_cor_E}) and (\ref{Eq:H_cor_E}), we can compute the total AM flux through arbitrary observation planes.



%

\end{document}